\documentclass[a4paper,fleqn]{cas-dc}

\usepackage[numbers]{natbib}
\usepackage{amssymb} 
\usepackage{pifont}  
\newcommand{\cmark}{\checkmark} 
\newcommand{\xmark}{$\times$} 

\def\tsc#1{\csdef{#1}{\textsc{\lowercase{#1}}\xspace}}
\tsc{WGM}
\tsc{QE}

\begin{document}
\let\WriteBookmarks\relax
\def\floatpagepagefraction{1}
\def\textpagefraction{.001}

\shorttitle{FGCARL for Medical Image Segmentation}    

\shortauthors{Xiaobing Yu, Jin Yang}  

\title [mode = title]{FGCARL: Fisher-Guided Continuously Adaptive Representation Learning for Medical Image Segmentation}  



%

\author[1,3]{Xiaobing Yu}
\cormark[1]

\author[2]{Jin Yang}
\cortext[1]{Xiaobing Yu and Jin Yang contributed equally to this work and are co-first authors}
\cormark[1]

\author[1]{Peijie Qiu}
\author[1]{Weiwei Ma}
\author[3]{Xiaofeng Liu}
\ead{xiaofeng.liu@yale.edu}



\affiliation[1]{organization={Mallinckrodt Institute of Radiology},
            addressline={Washington University School of Medicine in St. Louis}, 
            city={St. Louis},
            postcode={63110}, 
            state={MO},
            country={USA}}

\affiliation[2]{organization={Department of Radiation Oncology},
            addressline={Icahn School of Medicine at Mount Sinai}, 
            city={New York},
            postcode={10029}, 
            state={NY},
            country={USA}}


\affiliation[3]{organization={Departments of Radiology and Biomedical Imaging, and Biomedical Informatics and Data Science},
            addressline={Yale University}, 
            city={New Haven},
            postcode={06519}, 
            state={CT},
            country={USA}}


\begin{abstract}
Deep learning (DL) has achieved remarkable success in medical image segmentation by learning data-driven representations from large-scale annotated datasets. However, medical images are often collected independently across clinical sites, resulting in substantial domain shifts due to differences in imaging modalities, acquisition protocols, and intensity distributions. Meanwhile, segmentation classes may vary across sites due to differences in annotation standards, leading to class shifts. These evolving domains and class definitions can substantially degrade segmentation performance on newly encountered data, while sequential adaptation may cause catastrophic forgetting of knowledge acquired from previously encountered domains and classes. Existing incremental learning methods for medical image segmentation primarily address either domain shifts or class shifts, with these two settings largely investigated independently. This separation leaves a critical gap in developing a unified approach that can simultaneously adapt to evolving domains and newly introduced classes while retaining previously acquired knowledge. To address this challenge, we propose \textbf{Fisher-Guided Continuously Adaptive Representation Learning (FGCARL)}, a unified framework for domain- and class-incremental 3D volumetric medical image segmentation. FGCARL comprises two key components: \textbf{Mutual Adaptive Instance Representation (MAIR)}, which adaptively aligns feature distributions across sequentially encountered domains through bidirectional normalization and denormalization, and \textbf{Fisher-Guided Knowledge Retention (FGKR)}, which selectively preserves parameters important for previously learned domains and classes through Fisher-guided regularization. We conducted extensive experiments across six cross-institutional segmentation tasks using diverse segmentation architectures. Experimental results demonstrate that FGCARL consistently outperformed state-of-the-art continual learning methods, effectively adapting to domain and class shifts while mitigating catastrophic forgetting and maintaining strong segmentation performance on both previously learned and newly encountered domains.
 
\end{abstract}





\maketitle
\section{Introduction}
\begin{figure}
    \centering
    \includegraphics[width=\linewidth]{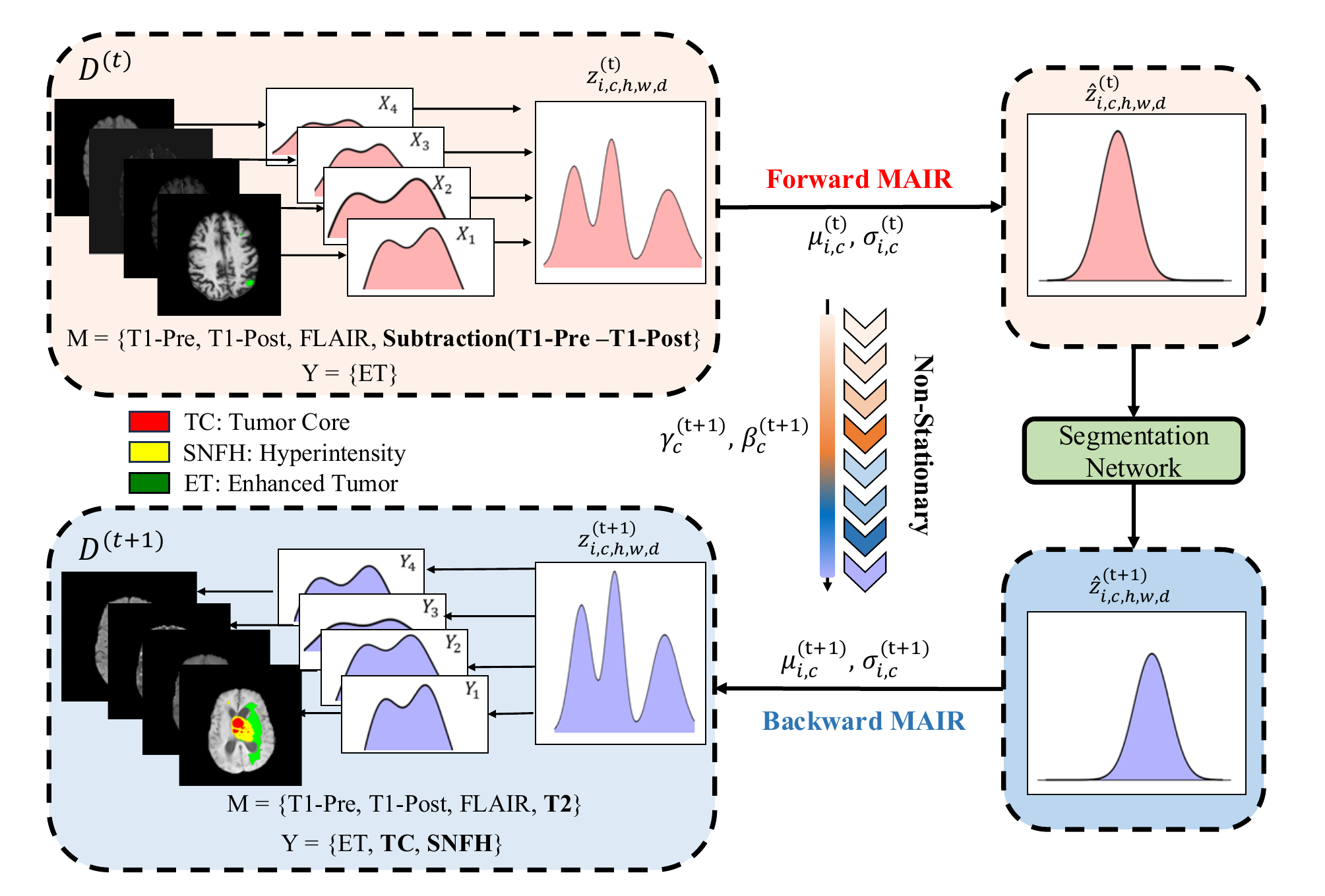} 
    \caption{A two-stage incremental learning framework for medical image segmentation. The segmentation networks were adapted from the UCSF-BMSR domain \(D^t\) to the BraTS-METS domain \(D^{t+1}\). The segmentation network was trained to segment ET from three modalities, and adapted to segment ET, TC, and SNFH from four modalities.}
    \label{fig1}
\end{figure}

Segmentation of anatomical structures and pathological regions in medical images plays a critical role in supporting diagnosis, prognosis, and treatment planning. It has markedly progressed by the development in deep learning (DL)-based methods \cite{ronneberger2015u,cciccek20163d,azad2024medical,yang2025dmc,yang2025d2}. DL models achieve impressive segmentation performance due to their ability to capture rich, data-driven features \cite{isensee2021nnu,zhou2023nnformer,YANG2026108837,qiu2026agileformer}. Typically, these models are trained on a specific domain to effectively learn features from domain-specific data. However, when these models are employed to make zero-shot predictions on new domains, they fail to capture domain-specific patterns, thus limiting their generalizability and applicability to unseen datasets or tasks \cite{ma2024segment,yang2025active}. Moreover, in clinical practice, DL models are generally developed for specific tasks. However, clinical datasets are often constructed and developed incrementally across multiple sites, each using different imaging modalities and acquisition protocols or parameters \cite{liu2021subtype,liu2022act,liu2023attentive,liu2023memory}. Additionally, the set of annotated anatomical structures may expand over time as new lesions or structures of interest are introduced, driven by the specific goals of the study, clinical needs, or newly available modalities \cite{ozdemir2018learn,liu2022learning}. Specifically, as illustrated in Figure \ref{fig1}, a segmentation model was trained to segment enhancing tumor (ET) from three modalities, including pre-contrast T1 weighted (T1w-pre), post-contrast T1-weighted (T1w-post), and Fluid-Attenuated Inversion Recovery (FLAIR), in the UCSF-BMSR dataset \cite{rudie2024university}. However, the model was subsequently adapted to segment ET and two additional structures, tumor core (TC) and surrounding non-enhancing FLAIR hyperintensity (SNFH), with the inclusion of additional T2-weighted (T2w) modality in the BraTS-METS 2023 dataset \cite{moawad2024brain}. This example highlights the challenges posed by domain shifts and incremental annotation in medical image segmentation.

To enable continual model adaptation to evolving clinical data, incremental learning (IL) frameworks have been increasingly explored in medical image segmentation. Existing approaches can be broadly categorized into two paradigms: \textit{domain-incremental learning} (DIL) and \textit{class-incremental learning} (CIL). DIL focuses on adapting segmentation models to sequentially encountered domains with different data distributions, such as those arising from changes in imaging modalities, acquisition protocols, or clinical sites \cite{you2022incremental,gonzalez2023lifelong,bera2023memory,liu2023incremental}. In contrast, CIL addresses the sequential introduction of new anatomical structures or pathological categories, enabling models to learn additional segmentation classes without retraining from scratch \cite{ozdemir2018learn,douillard2021plop,yu2022self,yang2022uncertainty,liu2022learning}. Although these approaches have advanced continual adaptation for medical image segmentation, several important challenges remain.

\textbf{First, existing methods predominantly consider domain and class shifts independently.} Most DIL and CIL approaches are specifically designed for either evolving data distributions or expanding label spaces, while real-world clinical deployment may involve both changes simultaneously. For example, a model may be sequentially deployed across institutions that use different imaging modalities while also requiring segmentation of newly introduced anatomical structures or lesion types. Although a few class-domain-incremental learning approaches have been explored in natural image recognition \cite{kundu2020class,park2024versatile}, their underlying assumptions and learning mechanisms do not directly translate to medical image segmentation. In particular, many natural-image approaches rely on image-level class representations, prototypes, or classification-oriented objectives, whereas medical image segmentation requires dense voxel-wise predictions and preservation of spatially structured anatomical representations. Consequently, a unified framework specifically designed for simultaneous domain and class shifts in volumetric medical image segmentation remains underexplored.

\textbf{Second, rehearsal-based strategies for mitigating catastrophic forgetting raise important challenges in clinical applications.} Rehearsal methods preserve knowledge from previous tasks by storing and replaying representative samples or exemplars \cite{gonzalez2023lifelong}. However, retaining patient imaging data for continual model training can introduce privacy, security, and data-governance concerns, particularly in multi-institutional clinical settings where patient data may be subject to strict access and retention policies \cite{li2022domain}. This motivates rehearsal-free approaches that can preserve previously acquired knowledge without retaining sensitive patient data.

\textbf{Third, existing adaptation mechanisms may be closely coupled to specific segmentation architectures and normalization strategies.} This issue becomes particularly relevant as modern medical image segmentation increasingly incorporates diverse architectures, including convolutional neural networks (CNNs) and vision transformers (ViTs). For example, ViT-based models commonly employ layer normalization (LN) \cite{ba2016layer}, which normalizes feature statistics within individual samples and may not explicitly account for domain-level distribution differences. Such normalization can therefore limit the model's ability to explicitly adapt its feature representations to substantial domain shifts. More broadly, architecture-specific adaptation mechanisms may hinder the development of a general continual learning strategy that can be consistently applied across heterogeneous segmentation architectures.

These limitations limit the applicability and generalizability of current IL methods on medical image segmentation, and highlight the need for a unified, rehearsal-free, and architecture-agnostic framework that can simultaneously accommodate evolving domains and class spaces while preserving previously acquired knowledge. Such a framework is particularly important for clinical scenarios in which segmentation models must continuously adapt to new institutions, imaging protocols, and anatomical or pathological targets without access to historical patient data.

To address these limitations, we propose \textbf{Fisher-Guided Continuously Adaptive Representation Learning (FGCARL)}, a unified framework for domain- and class-incremental learning in 3D volumetric medical image segmentation. FGCARL is designed to simultaneously accommodate evolving imaging domains, including differences in modalities, acquisition protocols, and institutional distributions, and expanding or changing segmentation label spaces, without requiring access to previously observed patient data. The framework consists of two complementary components: \textbf{Mutual Adaptive Instance Representation (MAIR)} and \textbf{Fisher-Guided Knowledge Retention (FGKR)}.

The \textbf{MAIR} module addresses the representation-level challenges arising from domain shifts. It performs bidirectional feature transformation through adaptive normalization and denormalization, enabling the model to establish consistent feature representations across previously encountered and newly observed domains. Specifically, MAIR normalizes feature representations during the forward propagation to reduce domain-specific distribution discrepancies and subsequently restores domain-relevant statistics through denormalization during backward propagation, allowing domain-specific information to be adaptively incorporated during optimization. Unlike methods that rely on storing historical samples or explicitly maintaining domain-specific representations, MAIR performs adaptation directly in the feature space and therefore does not require retention of raw patient data. Moreover, its feature-level formulation is decoupled from task-specific output spaces and sequence-length dependencies, facilitating integration with heterogeneous segmentation architectures, including both convolutional neural networks (CNNs) and vision transformers (ViTs).

The \textbf{FGKR} strategy addresses catastrophic forgetting during sequential adaptation. Rather than relying on rehearsal of historical exemplars, FGKR estimates the importance of model parameters to previously learned tasks using Fisher information and selectively constrains updates to parameters that carry critical historical knowledge. This parameter-level knowledge retention enables the model to acquire new domain- and class-specific knowledge while preserving information accumulated from previous tasks, without requiring storage or replay of sensitive patient data.

FGCARL therefore provides a unified solution to the three key challenges in medical image segmentation continual learning: simultaneous domain and class shifts, privacy constraints associated with rehearsal-based learning, and adaptation across heterogeneous segmentation architectures. Unlike conventional domain-specific normalization approaches \cite{chang2019domain,liu2020ms,du2023mdvit}, MAIR performs bidirectional statistical adaptation without requiring access to previously observed datasets. Furthermore, its non-parametric distribution matching formulation does not impose restrictive Gaussian assumptions on medical imaging feature distributions, allowing it to accommodate complex and potentially non-Gaussian representations. By combining representation-level domain adaptation with Fisher-guided parameter-level knowledge retention, FGCARL enables continual learning across domains with different imaging characteristics and evolving anatomical or pathological label spaces while mitigating catastrophic forgetting.

Extensive experiments across multiple cross-institutional segmentation tasks and heterogeneous segmentation architectures demonstrate that FGCARL consistently outperforms state-of-the-art incremental learning methods, validating its effectiveness, robustness, and generalizability for domain- and class-incremental medical image segmentation. Our contributions are summarized as follows:

\begin{itemize}
\item We propose \textbf{Fisher-Guided Continuously Adaptive Representation Learning}, a unified framework for domain- and class-incremental learning in 3D volumetric medical image segmentation. FGCARL simultaneously addresses evolving imaging domains and segmentation label spaces while supporting adaptation across heterogeneous segmentation architectures.

\item We introduce \textbf{Mutual Adaptive Instance Representation}, a feature-level adaptation module that performs bidirectional normalization and denormalization to align feature representations across sequentially encountered domains. MAIR enables effective adaptation to distribution shifts without requiring access to previously observed patient data and can be integrated with diverse segmentation architectures.

\item We develop \textbf{Fisher-Guided Knowledge Retention}, a rehearsal-free optimization strategy that uses Fisher information to selectively preserve parameters critical to previously learned domains and classes. FGKR mitigates catastrophic forgetting while allowing the model to acquire new knowledge during incremental training without storing historical patient data.

\item We conduct extensive cross-institutional domain- and class-incremental experiments across multiple medical image segmentation tasks and architectures. FGCARL consistently outperforms state-of-the-art incremental learning methods, demonstrating its effectiveness and generalizability across diverse segmentation settings.

\end{itemize}

\section{Related Works}
\subsection{Class Incremental Learning} 
CIL aims to sequentially incorporate newly introduced classes while preserving the segmentation performance of previously learned classes. Existing CIL approaches can be broadly categorized into four main paradigms. First, replay-based methods retain a subset of historical data or generate synthetic samples and replay them during incremental training to alleviate catastrophic forgetting \cite{bera2023memory,rebuffi2017icarl,hou2019learning,Tiwari2022CVPR,wang2022,gao2022r,li2024comprehensive,liu2022few,shin2017continual}. Although effective, storing or replaying historical medical images and annotations can introduce substantial privacy, security, and data-governance concerns in clinical applications. Second, architecture-based methods dynamically expand model capacity by introducing task-specific branches, sub-networks, or other architectural components for newly encountered classes \cite{fernando2017pathnet,hung2019compacting,kanakis2020reparameterizing,li2019learn,liu2021adaptive,zhang2022representation}. While these approaches can isolate newly learned knowledge from previously acquired representations, continual network expansion increases model complexity and may limit scalability and consistency across incremental tasks. Third, regularization-based methods constrain updates to parameters or representations that are important for previously learned classes, thereby preserving historical knowledge while adapting the model to new classes \cite{chaudhry2018riemannian,joseph2022energy,kirkpatrick2017overcoming,liu2023incremental,ozgun2020importance}. However, existing regularization strategies often rely on parameter-importance estimation or optimization mechanisms that are closely coupled with specific model architectures, which can limit their generalizability across heterogeneous segmentation backbones. Fourth, knowledge-distillation-based methods transfer knowledge from a previously trained model to the incrementally updated model by encouraging consistency between their predictions or intermediate representations, thereby reducing forgetting of previously learned classes \cite{bera2023memory,douillard2021plop,liu2022learning,park2024mitigating,tan2024semantically,wang2024incremental,yang2022uncertainty,yu2022self}. However, distillation-based approaches typically require maintaining a previous model and designing appropriate distillation objectives to preserve knowledge across changing label spaces, which can become challenging when newly introduced classes and previously learned classes have substantially different distributions or semantic relationships.
 
\subsection{Domain Incremental Learning} 
DIL focuses on sequentially adapting models to evolving data distributions, enabling segmentation models to accommodate changes in imaging modalities, acquisition protocols, and institutional characteristics. Similar to CIL, however, DIL is inherently susceptible to catastrophic forgetting, as adaptation to a newly encountered domain can overwrite knowledge acquired from previously observed domains. Early DIL approaches commonly employed exemplar-based rehearsal, retaining data from previous domains for joint training with the current domain \cite{you2022incremental}. To reduce the dependence on historical data, Li et al. proposed recovering information from previously encountered domains using an auxiliary generative model \cite{li2022domain}. However, generating realistic and sufficiently representative medical images solely from a trained model remains challenging \cite{yin2020dreaming} and may introduce additional privacy risks if sensitive patient information is inadvertently encoded or reproduced \cite{zhang2021unsupervised}. More recently, several studies have explored extending DIL toward more general continual learning settings. Nevertheless, existing approaches remain constrained by important assumptions that limit their applicability to medical image segmentation. Some methods are primarily developed for natural-image classification and therefore do not directly accommodate dense, voxel-wise prediction in 3D medical image segmentation \cite{kundu2020class,park2024versatile}. Other approaches are specifically designed for conventional convolutional architectures and may not readily generalize to transformer-based segmentation models \cite{gonzalez2023lifelong,liu2023incremental}. In addition, several methods rely on access to previously observed data during incremental adaptation \cite{gonzalez2020wrong,gonzalez2023lifelong}, which is undesirable in clinical settings because historical patient data may be inaccessible across institutions or subject to strict privacy and data-governance constraints. These limitations become particularly pronounced for high-dimensional, multimodal 3D medical images, where effective continual adaptation must preserve rich spatial and semantic representations while accommodating substantial distribution shifts. Normalization-based approaches have also been explored for domain adaptation by adjusting feature statistics to reduce distribution discrepancies across domains \cite{nam2018batch,mirza2022norm,bi2024learning}. However, conventional instance normalization and related approaches primarily focus on aligning representations toward the current target domain and do not explicitly preserve domain-specific knowledge acquired from previously encountered domains. Consequently, directly applying such approaches to DIL may improve adaptation to the current domain at the expense of historical knowledge. This motivates a continual representation adaptation mechanism that can not only adapt feature statistics to newly encountered domains but also retain and recover knowledge associated with previously learned domains without requiring access to historical patient data.

\section{Methods}
\subsection{Overview of FGCARL}
Let $\mathcal{D}^{(t)}$ denote the dataset observed at incremental stage $t$, consisting of $N_t$ 3D medical images $x_i \in \mathbb{R}^{H\times W\times D\times C}$ with segmentation annotations $y_i \in \mathbb{R}^{H\times W\times D}$ (where $H$, $W$, and $D$ denote the height, width, and depth of each imaging volume, respectively):
\begin{equation}
    \mathcal{D}^{(t)} = \{(x_i^{(t)},y_i^{(t)}) \mid y_i^{(t)} \in \{0,\ldots,K^{(t)}\}\}_{i=1}^{N_t},
\end{equation}
where $C$ represents the number of input channels. $K^{(t)}$ denotes the number of segmentation classes observed at stage $t$. The superscript $(t)$ indicates the incremental stage at which the corresponding image and annotation are acquired. At each incremental stage, the segmentation model is adapted from domain $t$ to a newly encountered domain $t+1$. In the proposed domain- and class-incremental setting, both the feature distributions and the label spaces may change across stages, i.e.,
\begin{align}
    P^{(t)}(X)        &\neq P^{(t+1)}(X), \\
    \mathcal{Y}^{(t)} &\neq \mathcal{Y}^{(t+1)},
\end{align}
where the former represents domain shift and the latter represents class shift. These changes introduce two complementary challenges: the model must acquire new domain- and class-specific knowledge while preserving knowledge acquired from previously encountered domains and anatomical structures.

\begin{figure*}
    \centering
    \includegraphics[width=\linewidth]{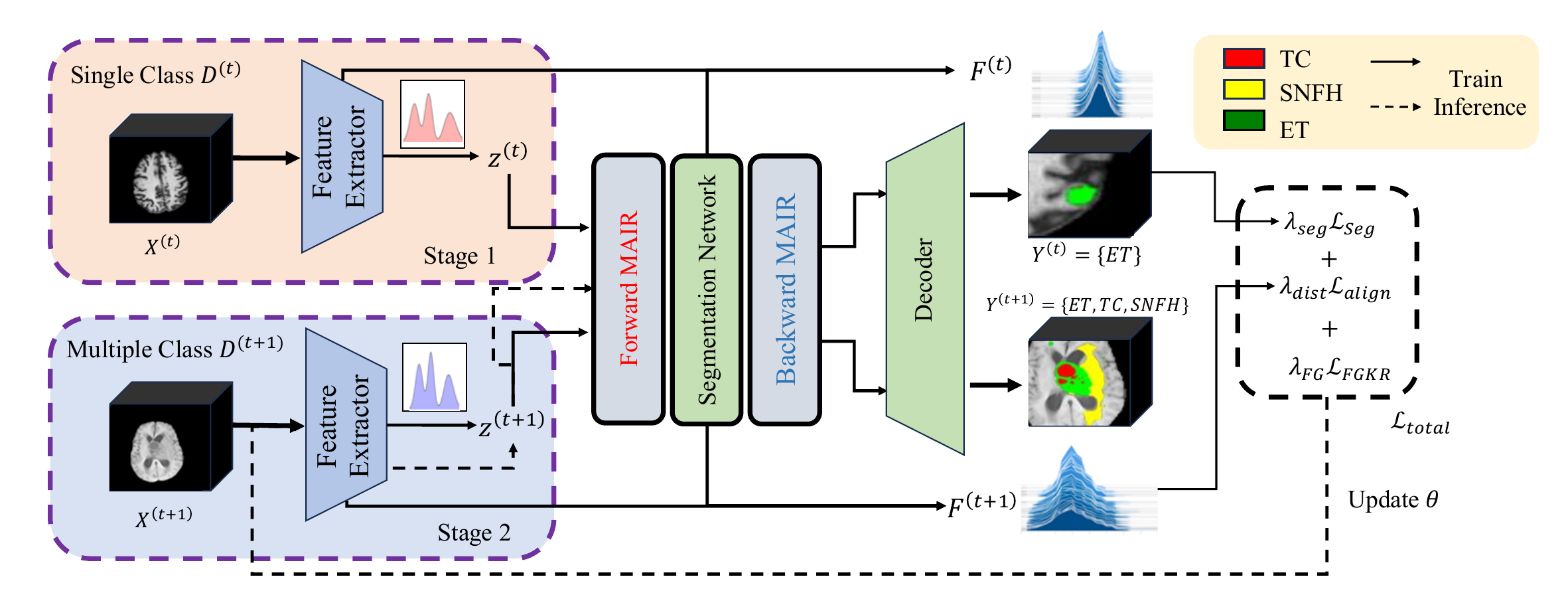}
    \caption{The overall architecture of FGCARL. FGCARL adapts the segmentation network from a single-class domain to a multi-class domain by employing MAIR and FGKR components. The input features $x^{(t)}$ and $x^{(t+1)}$ are extracted from the feature extractor. Subsequently, MAIR is employed to the segmentation network by incorporating Forward MAIR and Backward MAIR by aligning features $x^{(t)}$ to $x^{(t+1)}$ across domains and classes. A shared decoder is utilized to generate final segmentation results. FGKR is employed as an optimization mechanism by utilizing a loss function $L_{FGKR}$ which is calculated from the Fisher information matrix $F^{(t)}$ and $F^{(t+1)}$.}
    \label{fig:fgcarl}
\end{figure*}

To address these challenges, we propose \textbf{Fisher-Guided Continuously Adaptive Representation Learning}, a unified framework that jointly performs feature-level domain adaptation and parameter-level knowledge retention (Figure~\ref{fig:fgcarl}). FGCARL consists of two complementary components: \textbf{Mutual Adaptive Instance Representation} and \textbf{Fisher-Guided Knowledge Retention}. MAIR reduces domain-specific distributional discrepancies while retaining a connection to previously learned feature representations, whereas FGKR selectively constrains parameters that are important for previously acquired knowledge. Together, these components establish a stability--plasticity mechanism that enables the model to adapt to new domains and classes without requiring rehearsal of historical patient data.

\subsection{Mutual Adaptive Instance Representation}
Medical images acquired from different institutions may exhibit substantial distributional differences caused by imaging modalities, acquisition protocols, scanners, reconstruction procedures, and intensity characteristics. Such differences can propagate through the segmentation network and result in inconsistent intermediate feature representations. Directly adapting the model to the newly encountered domain may therefore cause previously learned feature statistics to be overwritten.

\begin{figure*}[t]
    \centering
    \includegraphics[width=0.9\linewidth]{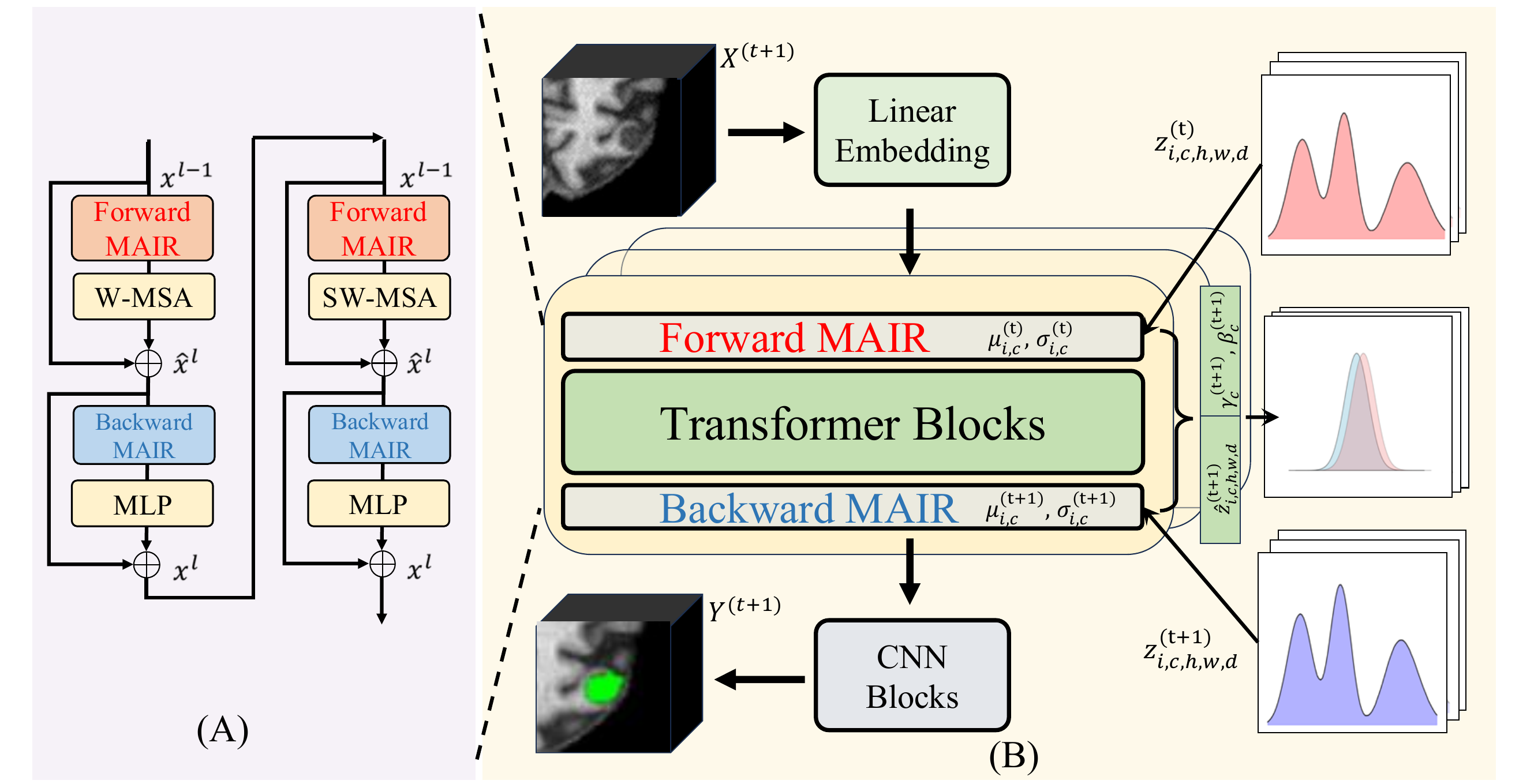}
    \caption{Transformer-based architecture incorporating MAIR for 3D medical image segmentation. (A) Detailed structure of a transformer block equipped with MAIR layers. (B) Overall architecture of the proposed MAIR-equipped transformer framework. Given an input 3D medical image $X^{(t+1)}$, a linear embedding layer first projects the image into feature embeddings, which are subsequently normalized by the Forward MAIR module before being processed by the transformer blocks. The resulting feature representations are then denormalized by the Backward MAIR module and fed into the CNN decoder blocks for hierarchical feature reconstruction. Finally, the decoder generates the segmentation prediction $Y^{(t+1)}$.}
    \label{fig:mair}
\end{figure*}

To address this issue, we propose \textbf{Mutual Adaptive Instance Representation (MAIR)} (Figure~\ref{fig:mair}), which separates domain-dependent feature statistics from the underlying feature structure and enables bidirectional adaptation between previously encountered and newly encountered domains. MAIR consists of two complementary operations: \textit{Forward MAIR} and \textit{Backward MAIR}. Forward MAIR normalizes the feature representation of the newly encountered domain and applies learnable affine transformations to retain the flexibility required for domain-specific adaptation. Backward MAIR subsequently reconstructs the representation using statistics associated with a previously learned domain. In this way, MAIR provides a mechanism for adapting to new feature distributions while maintaining statistical consistency with previously acquired representations.

Importantly, MAIR operates directly on intermediate feature representations and therefore does not require storing or replaying historical patient images or annotations. Its feature-level formulation also makes it applicable to different segmentation architectures, including convolutional neural networks (CNNs) and vision transformers (ViTs).

\subsubsection{Forward MAIR}
Given a feature embedding $ \mathbf{z}^{(t+1)} \in \mathbb{R}^{B\times C\times H\times W\times D}$ extracted from the newly encountered domain $t+1$, Forward MAIR first computes instance-specific statistics independently for each sample and channel. For the $i$-th instance and $c$-th channel, the mean is calculated over the spatial dimensions as
\begin{align}
    \mu_{i,c}^{(t+1)} = \frac{1}{HWD}\sum_{h,w,d}z_{i,c,h,w,d}^{(t+1)},
\end{align}
and the corresponding variance is
\begin{align}
    \left(\sigma_{i,c}^{(t+1)}\right)^2=\frac{1}{HWD}\sum_{h,w,d}\left(z_{i,c,h,w,d}^{(t+1)}-\mu_{i,c}^{(t+1)}\right)^2.
\end{align}
The feature representation is then normalized as
\begin{align}
    \tilde{z}_{i,c,h,w,d}^{(t+1)}=\frac{z_{i,c,h,w,d}^{(t+1)}-\mu_{i,c}^{(t+1)}}{\sqrt{\left(\sigma_{i,c}^{(t+1)}\right)^2+\epsilon}},
\end{align}
where $\epsilon$ is a small constant introduced for numerical stability. The normalization removes instance-specific shifts and scales from the feature representation, thereby reducing the direct influence of domain-specific statistics. However, completely eliminating these statistics may also discard information that is useful for representing the newly encountered domain. Therefore, we subsequently introduce learnable channel-wise affine transformations:
\begin{align}
    \hat{z}_{i,c,h,w,d}^{(t+1)}=\gamma_c^{(t+1)}\cdot\tilde{z}_{i,c,h,w,d}^{(t+1)}+\beta_c^{(t+1)},
\end{align}
where $\gamma_c^{(t+1)}$ and $\beta_c^{(t+1)}$ denote learnable scaling and shifting parameters, respectively. The scaling parameter $\gamma_c^{(t+1)}$ controls the magnitude of the normalized representation, whereas $\beta_c^{(t+1)}$ controls its location. Consequently, Forward MAIR does not simply eliminate domain-specific information; instead, it provides learnable degrees of freedom that allow the model to recover feature statistics beneficial for the newly encountered domain. This design balances domain invariance and domain adaptability.

\textbf{Statistical Stability of Forward MAIR.} Forward MAIR provides a controlled feature representation regardless of the original feature scale. Specifically, because the instance-wise mean is explicitly removed, the mean of the normalized features is
\begin{align}
    \nonumber
    \mathbb{E}[\tilde{z}_{i,c}^{(t+1)}] &= \mathbb{E}\left[\frac{z_{i,c}^{(t+1)}-\mu_{i,c}^{(t+1)}}{\sqrt{\left(\sigma_{i,c}^{(t+1)}\right)^2+\epsilon}}\right] \\
    &= \frac{\mathbb{E}[z_{i,c}^{(t+1)}-\mu_{i,c}^{(t+1)}]}{\sqrt{\left(\sigma_{i,c}^{(t+1)}\right)^2+\epsilon}} = 0.
\end{align}
Furthermore, the variance of the normalized feature is
\begin{align}
    \nonumber
    \mathrm{Var}[\tilde{z}_{i,c}^{(t+1)}] &= \frac{\mathrm{Var}\left[z_{i,c,h,w,d}^{(t+1)}-\mu_{i,c}^{(t+1)}\right]}{\left(\sigma_{i,c}^{(t+1)}\right)^2+\epsilon} \\
    & = \frac{\left(\sigma_{i,c}^{(t+1)}\right)^2}{\left(\sigma_{i,c}^{(t+1)}\right)^2+\epsilon} \leq 1.
\end{align}
Thus, Forward MAIR maps the feature representation into a controlled statistical range before subsequent network operations. This property is particularly useful under domain shifts, where changes in imaging conditions can produce substantial variations in feature magnitude and dispersion. The subsequent learnable affine transformation restores the flexibility necessary for domain-specific adaptation.

\subsubsection{Backward MAIR}
Although Forward MAIR reduces the influence of domain-specific statistics, normalization alone does not explicitly preserve the statistical characteristics of previously encountered domains. Consequently, continuous adaptation exclusively toward the newly encountered domain may progressively move the feature space away from representations that supported previous segmentation performance.

To address this issue, Backward MAIR reintroduces feature statistics associated with a previously learned domain. Let $\mu_c^{(t)}$ and $\sigma_c^{(t)}$ denote the reference mean and standard deviation associated with domain $t$. Given the affine-transformed representation $\hat{z}^{(t+1)}$, Backward MAIR first removes the current-domain affine transformation and then restores the reference-domain statistics:
\begin{align}
    z_{i,c,\mathrm{back}}^{(t)} =\sigma_c^{(t)}\cdot\left(\frac{\hat{z}_{i,c}^{(t+1)}-\beta_c^{(t+1)}}{\gamma_c^{(t+1)}}\right)+\mu_c^{(t)},
\end{align}
where $\gamma_c^{(t+1)}\neq0$. Because $\hat{z}_{i,c}^{(t+1)}=\gamma_c^{(t+1)}\cdot\tilde{z}_{i,c}^{(t+1)}+\beta_c^{(t+1)}$, the Backward MAIR transformation can equivalently be expressed as
\begin{align}
    z_{i,c,\mathrm{back}}^{(t)}=\sigma_c^{(t)}\cdot\tilde{z}_{i,c}^{(t+1)}+\mu_c^{(t)}.
\end{align}
This formulation illustrates the central mechanism of MAIR: the normalized representation retains the structural and semantic information extracted from the newly encountered input, while its feature statistics are mapped toward those of the previously learned domain. Importantly, this operation requires only compact feature statistics rather than historical patient images or annotations, thereby enabling rehearsal-free continual adaptation.

\textbf{Cross-Domain Alignment by Backward MAIR.} The statistical behavior of Backward MAIR can be characterized directly from the properties of the normalized representation. Since $\mathbb{E}\left[\tilde{z}_{i,c}^{(t+1)}\right]=0$, the mean of the backward-transformed representation satisfies
\begin{align}
    \mathbb{E}\left[z_{i,c,\mathrm{back}}^{(t)}\right] = \sigma_c^{(t)}\cdot\mathbb{E}\left[\tilde{z}_{i,c}^{(t+1)}\right]+\mu_c^{(t)} =\mu_c^{(t)}.
\end{align}
Similarly, its variance is given by
\begin{align}
    \nonumber
    \mathrm{Var}\left[z_{i,c,\mathrm{back}}^{(t)}\right] &= \left(\sigma_c^{(t)}\right)^2\cdot\mathrm{Var}\left[\tilde{z}_{i,c}^{(t+1)}\right] \\
    \nonumber
    &=\left(\sigma_c^{(t)}\right)^2\cdot\frac{\left(\sigma_{i,c}^{(t+1)}\right)^2}{\left(\sigma_{i,c}^{(t+1)}\right)^2+\epsilon} \\
    &\approx\left(\sigma_c^{(t)}\right)^2,
\end{align}
where the approximation becomes increasingly accurate as $\epsilon$ becomes small relative to the feature variance.

Therefore, Backward MAIR approximately restores the first two moments of the reference domain while preserving the normalized feature structure generated from the newly encountered domain. This bidirectional mechanism allows FGCARL to simultaneously accommodate distributional changes and maintain statistical continuity with previously learned domains.

\subsection{Non-Gaussian Distribution Alignment}
Although instance normalization and affine transformations regulate the first two moments of feature distributions, medical imaging features may exhibit substantial non-Gaussian characteristics, including skewness, heavy tails, and multi-modality, due to heterogeneous tissue composition, pathological variations, imaging artifacts, and differences in acquisition protocols. Consequently, aligning only the mean and variance may not sufficiently characterize distributional discrepancies across domains.

To address this limitation, FGCARL incorporates higher-order moment matching as a non-parametric distribution alignment mechanism. Rather than imposing a specific parametric distribution on the learned features, the proposed formulation progressively constrains the statistical characteristics of the feature distributions through multiple moments.

\subsubsection{Moment-Based Distribution Matching.}
Let $\mathcal{M}_{m,c}(\mathbf{z})$ denote the empirical $m$-th moment of the feature distribution for channel $c$, defined as
\begin{align}
    \mathcal{M}_{m,c}(\mathbf{z})=\frac{1}{N}\sum_{n=1}^{N}z_{n,c}^{m},
\end{align}
where $N$ denotes the number of feature observations used to estimate the moment. Given the normalized representations from the incoming domain $t+1$ and the reference domain $t$, we define the higher-order distribution alignment loss as
\begin{align}
    \mathcal{L}_{\mathrm{mom}}^{(t+1)}=\sum_{m=1}^{M}\lambda_m\sum_{c=1}^{C}\left[\mathcal{M}_{m,c}\left(\tilde{\mathbf{z}}^{(t+1)}\right)-\mathcal{M}_{m,c}\left(\tilde{\mathbf{z}}^{(t)}\right)\right]^2,
\end{align}
where $M$ denotes the maximum matched moment order and $\lambda_m$ controls the contribution of the $m$-th moment.

The first and second moments characterize feature location and dispersion, respectively, while higher-order moments provide additional information about the shape of the feature distribution. Thus, jointly matching multiple moments imposes progressively richer statistical constraints without requiring an explicit parametric assumption about the underlying feature distribution.

\subsubsection{Interpretation of Higher-Order Moment Matching}
Moment matching can be interpreted directly in terms of expectations of polynomial functions. Suppose two distributions $P$ and $Q$ have identical moments up to order $M$, such that
\begin{align}
    \mathbb{E}_{P}[X^m]=\mathbb{E}_{Q}[Y^m], \qquad m=0,\ldots,M.
\end{align}
For any polynomial $p(x)=\sum_{m=0}^{M}a_m x^m$ of degree at most $M$, it follows that
\begin{align}
    \nonumber
    \mathbb{E}_{P}[p(X)] &= \sum_{m=0}^{M}a_m\mathbb{E}_{P}[X^m] \\
    &= \sum_{m=0}^{M}a_m\mathbb{E}_{Q}[Y^m] = \mathbb{E}_{Q}[p(Y)].
\end{align}
Therefore, increasing the number of matched moments constrains the two feature distributions over an increasingly rich family of polynomial statistics. Importantly, matching a finite number of moments does not imply equality of the underlying distributions. Instead, it provides a practical approximation for reducing higher-order distributional discrepancies, while equality of all moments can uniquely characterize a distribution under appropriate moment-determinacy conditions.

\subsection{Embedding-Level Distribution Consistency}
While higher-order moment matching provides a flexible mechanism for reducing statistical discrepancies between feature distributions, it does not directly measure the overall discrepancy between the resulting embeddings. In particular, finite-order moment matching constrains only a finite set of statistical characteristics and may leave residual representation drift during incremental adaptation.

To explicitly regularize this residual drift, we introduce an embedding-level consistency loss based on the squared Wasserstein-2 ($W_2^2$) distance \cite{mallasto2017learning}. This loss serves a complementary role to higher-order moment matching: the latter focuses on progressively richer distributional statistics, whereas the Wasserstein term directly constrains the distance between the incoming representation and the representation realigned toward the previously learned domain.

Let $\tilde{\mathbf{z}}^{(t+1)}$ denote the normalized embedding obtained from Forward MAIR for the incoming domain $t+1$, and let $\hat{\mathbf{z}}_{\mathrm{back}}^{(t)}$ denote the corresponding embedding after Backward MAIR, which is realigned toward the statistics of the previously learned domain $t$. We approximate their channel-wise embedding distributions as Gaussian distributions:
\begin{align}
    \tilde{\mathbf{z}}^{(t+1)} &\sim \mathcal{N}\left(\boldsymbol{\mu}^{(t+1)}, \boldsymbol{\Sigma}^{(t+1)}\right), \\
    \hat{\mathbf{z}}_{\mathrm{back}}^{(t)} &\sim \mathcal{N}\left(\boldsymbol{\mu}^{(t)}, \boldsymbol{\Sigma}^{(t)}\right),
\end{align}
where $\boldsymbol{\mu}^{(t)}$ and $\boldsymbol{\mu}^{(t+1)}$ denote the corresponding channel-wise means, and $\boldsymbol{\Sigma}^{(t)}$ and $\boldsymbol{\Sigma}^{(t+1)}$ denote the covariance matrices. This Gaussian approximation is introduced solely to obtain a tractable discrepancy measure and does not imply that the underlying learned embeddings are inherently Gaussian.

For two multivariate Gaussian distributions, the squared Wasserstein-2 distance has the closed-form expression
\begin{align}
    \nonumber
    & W_2^2\left(\tilde{\mathbf{z}}^{(t+1)}, \hat{\mathbf{z}}_{\mathrm{back}}^{(t)}\right) \\
    \nonumber
    &= 
    W_2^2\left(\mathcal{N}(\boldsymbol{\mu}^{(t+1)},\boldsymbol{\Sigma}^{(t+1)}), \mathcal{N}(\boldsymbol{\mu}^{(t)},\boldsymbol{\Sigma}^{(t)})\right) \\
    &\tiny =\left|\boldsymbol{\mu}^{(t+1)}-\boldsymbol{\mu}^{(t)}\right|_2^2 +\operatorname{Tr}\left(\boldsymbol{\Sigma}^{(t+1)}+\boldsymbol{\Sigma}^{(t)}-2\left[(\boldsymbol{\Sigma}^{(t)})^{1/2}\boldsymbol{\Sigma}^{(t+1)}(\boldsymbol{\Sigma}^{(t)})^{1/2}\right]^{1/2}\right).
    \label{eq}
\end{align}
where $\operatorname{Tr}(\cdot)$ denotes the trace operator. In our implementation, the covariance matrices are represented using channel-wise variances and are therefore diagonal. Under this assumption, Eq. \ref{eq} is reduced to
\begin{equation}
    W_2^2\left(\left|\boldsymbol{\mu}^{(t+1)}-\boldsymbol{\mu}^{(t)}\right|_2^2+\left|\boldsymbol{\sigma}^{(t+1)}-\boldsymbol{\sigma}^{(t)}\right|_2^2\right),
\end{equation}
where $\boldsymbol{\sigma}^{(t)}$ and $\boldsymbol{\sigma}^{(t+1)}$ contain the channel-wise standard deviations. This formulation provides a computationally efficient measure of both location and scale discrepancies between the two embedding distributions.

Accordingly, the embedding consistency loss is defined as
\begin{equation}
    L_{\mathrm{align}}^{(t+1)}=W_2^2\left(\tilde{\mathbf{z}}^{(t+1)},\hat{\mathbf{z}}_{\mathrm{back}}^{(t)}\right).
\end{equation}
Minimizing $L_{\mathrm{align}}^{(t+1)}$ encourages the incoming-domain embedding to remain close to the representation that has been realigned toward the previously learned domain. Thus, the Wasserstein term provides an explicit representation-level constraint that complements the higher-order moment alignment.

The proposed alignment loss complements the two stages of MAIR by explicitly constraining the residual distributional discrepancy between incoming and previously learned representations. Forward MAIR first suppresses instance-specific intensity and feature-scale variations in the incoming domain, thereby providing a more stable representation for incremental adaptation. Backward MAIR subsequently restores the statistical characteristics associated with the previously learned domain, helping preserve historical knowledge while reducing representation drift. Building upon these two operations, the proposed alignment loss further regularizes the residual discrepancy between the incoming and historically aligned representations during optimization.

The proposed framework is particularly relevant to medical image segmentation, where domain shifts can arise from differences in imaging modality, acquisition protocols, scanner characteristics, and intensity distributions. Importantly, the alignment is performed directly in the learned embedding space and does not require storing or replaying historical medical images. Thus, the proposed approach provides an additional mechanism for controlling incremental representation drift while avoiding direct access to previously collected patient data.

The two alignment components within the framework play distinct but complementary roles. Higher-order moment matching characterizes and reduces discrepancies in the statistical shape of feature distributions without imposing an explicit parametric distributional assumption, making it suitable for capturing non-Gaussian characteristics of medical image representations. In contrast, Wasserstein alignment explicitly regularizes the residual discrepancy between the incoming and historical embeddings, providing a direct consistency constraint during optimization. These two mechanisms therefore operate at complementary levels: moment matching focuses on the statistical structure of the feature distributions, whereas Wasserstein alignment constrains the overall embedding-level discrepancy.

Overall, the combination of Forward and Backward MAIR with the two alignment mechanisms establishes a hierarchical adaptation strategy: \emph{forward normalization for representation stability, backward realignment for historical knowledge retention, higher-order moment matching for non-Gaussian distribution alignment, and Wasserstein regularization for explicit embedding consistency}. This design enables the model to simultaneously mitigate instance-level variation, preserve previously learned statistical characteristics, and constrain residual representation drift during incremental adaptation.

\subsection{Fisher-Guided Knowledge Retention}
While MAIR addresses feature-distribution shifts, feature alignment alone cannot guarantee the preservation of previously acquired segmentation knowledge. During incremental optimization, parameters that are important for previously learned domains or classes may be substantially modified to accommodate new data, resulting in catastrophic forgetting.

We therefore propose \textbf{Fisher-Guided Knowledge Retention (FGKR)}, which identifies parameters that are important for previously acquired knowledge and selectively constrains their updates. The underlying principle is to preserve parameters that have a large impact on the previous task while allowing less important parameters to remain flexible for learning new domains and classes.

\subsubsection{Fisher Information} 
Let $\theta^*$ denote the parameters obtained after learning domain $t$, and let $\hat{\theta}$ denote the parameters after adapting the model to domain $t+1$ without knowledge-retention regularization. A second-order Taylor expansion of the previous-domain loss around $\theta^*$ gives
\begin{align}
    \nonumber
    L^{(t)}(\hat{\theta}) &= L^{(t)}(\theta^*) + \nabla L^{(t)}(\theta^*)^{T} (\hat{\theta}-\theta^*) \\
    &+\frac{1}{2}(\hat{\theta}-\theta^*)^{T}H^{(t)}(\hat{\theta}-\theta^*) + O\left(|\hat{\theta}-\theta^*|^3\right),
\end{align}
where $H^{(t)}$ denotes the Hessian of the previous-domain loss. Assuming that $\theta^*$ is a local optimum, the first-order term is approximately zero. Under standard regularity conditions for likelihood-based models, the Fisher information provides a commonly used approximation to the local curvature of the loss:
\begin{align}
    H^{(t)}\approx F^{(t)}.
\end{align}
Using a diagonal approximation of the Fisher information matrix, the expected increase in the previous-domain loss can therefore be approximated as
\begin{align}
    L^{(t)}(\hat{\theta}) - L^{(t)}(\theta^*) \approx \frac{1}{2} \sum_j F_j^{(t)} \left(\hat{\theta}_j-\theta_j^*\right)^2.
\end{align}
The diagonal Fisher information for parameter $\theta_j$ can be estimated empirically as
\begin{align}
    \hat{F}_j^{(t)}=\frac{1}{|\mathcal{D}^{(t)}|}\sum_{(x,y)\in\mathcal{D}^{(t)}}\left[\frac{\partial\log p(y|x;\theta^*)}{\partial\theta_j}\right]^2.
\end{align}
A larger Fisher value indicates that perturbations to the corresponding parameter are associated with greater sensitivity of the previous-task objective. Such parameters are therefore considered more important for preserving previously acquired knowledge.

\subsubsection{Fisher-Guided Knowledge Retention}
Based on the estimated parameter importance, FGKR introduces the following Fisher-weighted regularization term:
\begin{align}
    \mathcal{L}_{\mathrm{FGKR}}^{(t+1)}=\frac{1}{2}\sum_j\hat{F}_j^{(t)}\left(\theta_j-\theta_j^*\right)^2.
\end{align}
The gradient of this regularization term with respect to an individual parameter is
\begin{align}
    \frac{\partial\mathcal{L}_{\mathrm{FGKR}}^{(t+1)}}{\partial\theta_j}=\hat{F}_j^{(t)}\left(\theta_j-\theta_j^*\right).
\end{align}
Therefore, parameters with large Fisher information receive stronger constraints and remain closer to their previously learned values, whereas parameters with small Fisher information can undergo larger updates to accommodate newly encountered domains and classes. FGKR consequently establishes an adaptive stability--plasticity trade-off rather than uniformly restricting all model parameters.

For sequential learning across multiple domains and classes, the Fisher information and corresponding reference parameters can be accumulated across incremental stages according to the implementation of the proposed framework. This allows historical knowledge to be retained throughout successive domain and class transitions without requiring replay of the corresponding historical patient data.

Unlike rehearsal-based approaches, FGKR preserves historical knowledge through parameter-level constraints rather than storing previous images or annotations. This property is particularly advantageous in clinical applications where historical patient data may be inaccessible because of privacy, security, or institutional data-governance constraints.

\subsection{Unified Incremental Learning Objective Function}
FGCARL jointly optimizes segmentation, distribution alignment, and knowledge retention at each incremental stage. The overall objective is defined as
\begin{align}
    \mathcal{L}_{\mathrm{total}}^{(t+1)}=\lambda_{\mathrm{seg}}\mathcal{L}_{\mathrm{seg}}^{(t+1)}+\lambda_{\mathrm{dist}}\mathcal{L}_{\mathrm{align}}^{(t+1)}+\lambda_{\mathrm{FG}}\mathcal{L}_{\mathrm{FGKR}}^{(t+1)},
\end{align}
where $\mathcal{L}_{\mathrm{seg}}^{(t+1)}$ denotes the segmentation loss on the newly encountered domain and classes, $\mathcal{L}_{\mathrm{align}}^{(t+1)}$ encourages consistency between feature distributions across incremental domains via Wasserstein-2 distance, and $\mathcal{L}_{\mathrm{FGKR}}^{(t+1)}$ preserves parameters that are important for previously acquired knowledge. The coefficients $\lambda_{\mathrm{dist}}$ and $\lambda_{\mathrm{FGKR}}$ control the contributions of the two auxiliary objectives.

The three terms play complementary roles. The segmentation objective provides the plasticity required to acquire knowledge from newly introduced domains and anatomical classes. MAIR reduces domain-specific distributional discrepancies and maintains statistical continuity with previously encountered domains. FGKR constrains updates to parameters that are important for previously acquired knowledge and thereby improves model stability. Through their joint optimization, FGCARL balances adaptation and retention, enabling a segmentation model to continuously learn from both evolving imaging domains and newly introduced anatomical classes without requiring rehearsal of historical patient data.

\section{Experiments}
\subsection{Datasets}
We evaluated FGCARL on six publicly available datasets covering diverse volumetric medical image segmentation tasks. These datasets span multiple imaging modalities, anatomical regions, disease types, acquisition settings, and segmentation targets, providing a comprehensive testbed for evaluating domain- and class-incremental learning.

\textbf{UCSF-BMSR.} The UCSF Brain Metastasis Segmentation and Registration (UCSF-BMSR) dataset contains 560 brain MR scans acquired using three modalities: T1-weighted pre-contrast (T1w-pre), T1-weighted post-contrast (T1w-post), and FLAIR. It provides voxel-wise annotations for 5,136 brain metastases and supports volumetric segmentation of metastatic lesions \cite{rudie2024university}.

\textbf{BrainMetShare.} BrainMetShare is a multi-institutional dataset comprising 156 whole-brain MR scans with four modalities: T1w-pre, T1w-post, T2-weighted (T2w), and FLAIR. Each scan includes voxel-wise annotations of brain metastases. Its multi-institutional nature introduces substantial acquisition and distribution variability, making it suitable for evaluating domain-incremental learning \cite{grovik2020deep}.

\textbf{BraTS-METS 2023.} The BraTS-METS 2023 dataset contains 1,303 multi-parametric brain MR scans with T1w-pre, T1w-post, T2w, and FLAIR modalities. It provides voxel-level annotations for three tumor regions: enhancing tumor (ET), tumor core (TC), and surrounding non-enhancing FLAIR hyperintensity (SNFH), providing a challenging setting for class-incremental segmentation \cite{moawad2024brain}.

\textbf{Multi-Center, Multi-Vendor and Multi-Disease Cardiac Segmentation (M$\&$Ms).} The M$\&$Ms dataset contains 375 cardiac MR scans acquired from multiple institutions using four MRI vendors, with 125, 125, 75, and 50 scans from the respective vendors. Voxel-wise annotations are provided for the left ventricle (LV), right ventricle (RV), and left ventricular myocardium (MYO). The multi-center and multi-vendor characteristics introduce substantial domain variability for evaluating domain-incremental learning \cite{campello2021multi}.

\textbf{Beyond the Cranial Vault (BTCV).} The BTCV dataset contains 50 abdominal CT volumes with voxel-wise annotations for 13 anatomical structures. The diversity of target organs and their anatomical characteristics provides a challenging benchmark for class-incremental segmentation \cite{landman2015miccai}.

\textbf{Liver Tumor Segmentation (LiTS).} The LiTS dataset contains 131 abdominal CT volumes with voxel-wise annotations for the liver and liver tumors. The substantial variation in tumor size, morphology, and appearance provides an additional challenging benchmark for incremental segmentation \cite{bilic2023liver}.

Collectively, these datasets enable evaluation of FGCARL across heterogeneous MR and CT imaging scenarios, multiple anatomical regions, and diverse segmentation targets, allowing us to assess its robustness to both domain and class shifts.

\subsection{Implementation details}
All experiments were implemented in PyTorch and conducted on a single NVIDIA Tesla A100 PCIe GPU with 40 GB of memory. We employed Swin UNETR as the vision transformer (ViT)-based backbone and nnU-Net as the convolutional neural network (CNN)-based backbone \cite{hatamizadeh2021swin,isensee2021nnu}. Both architectures were evaluated under the same incremental learning settings to assess the architecture-agnostic capability of FGCARL. The models were optimized using AdamW with an initial learning rate of $1\times10^{-4}$ and a weight decay of $1\times10^{-5}$. We adopted a linear warm-up followed by cosine annealing learning-rate scheduler, with a warm-up period of 50 epochs. All models were trained for 300 epochs with a batch size of 2.

Raw volumetric images were preprocessed to 3D patches of size $96\times96\times96$ and standardized using Z-score normalization. During training, patches were augmented using random flipping, random $90^\circ$ rotations, and random intensity scaling and shifting. Each dataset was divided into training and test sets at a ratio of $70\%:30\%$. Unless otherwise specified, all compared methods used the same data splits, preprocessing procedures, augmentation strategies, and training settings to ensure a fair comparison.

Segmentation performance was evaluated using the Dice Similarity Coefficient (DSC), reported as a percentage. For multi-class segmentation tasks, the DSC was calculated for each foreground class and then averaged across the target classes.

\subsection{Comparison with state-of-the-art methods}
We compared FGCARL with various state-of-the-art (SOTA) CIL and DIL methods for 3D multi-modal medical image segmentation. The methods include replay-, regularization-, architectural-, and knowledge-distillation-based approaches, covering both CNN- and ViT-based segmentation architectures. We considered two experimental stages to separately assess the segmentation capability before incremental adaptation and the ability to simultaneously accommodate domain and class shifts. In \textbf{Stage 1}, segmentation models were trained independently on the UCSF-BMSR and BraTS-METS 2023 datasets. These results provided a reference for the achievable segmentation performance when no incremental adaptation was implemented. In \textbf{Stage 2}, models were initially trained on UCSF-BMSR to segment ET using three modalities, including T1w-pre, T1w-post, and FLAIR, and were subsequently adapted to the BraTS-METS 2023 domain, where they were required to segment ET, TC, and SNFH using four modalities, including T1w-pre, T1w-post, T2w, and FLAIR. Therefore, this setting simultaneously introduced substantial \textit{domain shifts} caused by differences in data distributions and modality configurations and \textit{class shifts} caused by the introduction of new segmentation targets. To accommodate the different numbers of input modalities between the source and target domains, we additionally constructed a subtraction modality by computing the difference between T1w-pre and T1w-post images, thereby maintaining a consistent number of input channels across the two domains.

\begin{table*}[t!]
\centering
\caption{Performance comparison of FGCARL with SOTA class-incremental learning (CIL) and domain-incremental learning (DIL) methods under simultaneous domain and class shifts for 3D multi-modal medical image segmentation. $\dag$ and $*$ indicate CIL and DIL methods, respectively. “3D” and “ViT” indicate whether the method supports 3D segmentation and vision transformer (ViT)-based architectures, respectively. “3D” and “ViT” indicate whether the method supports 3D segmentation and vision transformer (ViT)-based architectures, respectively. DSC ($\%$) is reported for each segmentation class and as mean $\pm$ standard deviation. \textbf{Bold} and \underline{underline} indicate the best and the second best results, respectively.}
\resizebox{1.0\textwidth}{!}{
\begin{tabular}{@{}l|c|c|c|c|cccc|c|cccccc@{}}
\toprule
& CIL$\dag$& & & \multicolumn{5}{c|}{No Incremental Learning (Stage 1)} & \multicolumn{5}{c}{UCSF-BMSR $\rightarrow$ BraTS-METS 2023 (Stage 2)} \\ 
\cline{5-14}
\textbf{Methods} & or & 3D  & ViT & UCSF-BMSR & \multicolumn{4}{c|}{BraTS-METS 2023} & UCSF-BMSR &\multicolumn{4}{c}{BraTS-METS 2023} \\ 
\cline{5-14}
& DIL$*$& &  & ET & TC & SNFH & ET & Mean$\pm$SD & ET & TC & SNFH & ET & Mean$\pm$SD \\
\midrule
PLOP~~\cite{douillard2021plop}         & $\dag$     & \xmark & \xmark & 49.56 & 37.58 & 33.76 & 30.85 & 34.06 $ \pm$ 2.76 & 10.39 & 38.66 & 35.47 & 31.98 & 35.37 $\pm$ 2.73 \\
MargExcIL~~\cite{liu2022learning}      & $\dag$     & \xmark & \xmark & 51.75 & 41.41 & 40.47 & 48.59 &  43.49 $\pm$ 3.63 & 20.37 & 42.17 & 43.98 & 45.52 & 43.89 $\pm$ 1.37 \\
UCD~~\cite{yang2022uncertainty}        & $\dag$     & \xmark & \xmark & 56.37 & 42.15 & 45.72 & 41.66 & 43.18 $\pm$ 1.81 & 27.86 & 41.41 & 47.56 & 42.47 & 43.81 $\pm$ 2.68 \\
EWC~~\cite{kirkpatrick2017overcoming}  & $\dag$~$*$ & \xmark & \xmark & 55.51 & 41.96 & 43.65 & 47.18 & 47.08 $\pm$ 1.87 & 21.95 & 42.48 & 44.62 & 46.79 & 44.63 $\pm$ 1.76\\
HSI~~\cite{liu2022learning}            & $\dag$~$*$ & \xmark & \xmark & 65.46 & 52.51 & 55.28 & 51.91 & 56.29 $\pm$ 1.47 & 36.98 & 51.39 & 57.67 & 47.75 & 52.27 $\pm$ 4.10 \\
2DHIL~~\cite{imran2024two}             & $\dag$     & \xmark & \cmark & 53.17 & 47.27 & 45.89 & 50.78 & 47.98 $\pm$ 2.06 & 28.11 & 48.31 & 47.22 & 44.59 & 46.71 $\pm$ 1.56 \\
ACLSeg~~\cite{elskhawy2020continual}   & $\dag$     & \cmark & \xmark & 60.79 & 47.75 & 50.56 & 51.82 & 52.73 $\pm$ 1.70 & 31.41 & 47.45 & 53.97 & 49.76 & 50.39 $\pm$ 2.70 \\
NeST~~\cite{xie2024early}              & $\dag$     & \xmark & \cmark & 56.18 & 42.01 & 40.53 & 41.75 & 45.12 $\pm$ 0.65 & 34.79 & 42.97 & 41.73 & 37.77 &  40.82 $\pm$ 2.22 \\
Incrementer~~\cite{shang2023incrementer}& $\dag$    & \xmark & \cmark & 54.37 & 50.10 & 47.78 & 51.54 & 50.95 $\pm$ 1.55 & 28.57 & 51.54 & 50.77 & 39.85 & 47.39 $\pm$ 3.03 \\
MBS~~\cite{park2024mitigating}         & $\dag$    & \xmark & \cmark & 57.96 & 54.94 & 53.15 & 57.59 & 55.91 $\pm$ 1.82 & 31.91 & 55.82 & 54.67 & 52.61 & 54.37 $\pm$ 1.33 \\
SSIST~~\cite{tan2024semantically}      & $\dag$    & \xmark & \cmark & 59.81 & 60.33 & 62.15 & 50.44 & 58.18 $\pm$ 5.15 & 35.43 & 61.47 & 63.39 & 47.81 & 57.56 $\pm$ 6.94 \\
InNSeg~~\cite{wang2024incremental}     & $\dag$    & \xmark & \xmark  & 57.64 & 45.79 & 44.37 & 49.89 & 49.42 $\pm$ 2.34 & 35.87 & 43.71 & 46.58 & 45.89 & 45.94 $\pm$ 1.18 \\
CLAMTS~~\cite{zhang2023continual}      & $\dag$    & \cmark & \cmark  & 67.72 & \underline{66.42} & \underline{62.91} & 53.04 & \underline{60.79} $\pm$ 5.66  & 8.35 & \textbf{67.29} & \underline{64.57} & 50.95 & 60.94 $\pm$ 7.15 \\
Memory Replay~~\cite{bera2023memory}   & $\dag$~$*$ & \cmark & \xmark & 60.45 & 47.48 & 54.47 & 50.31 & 50.75 $\pm$ 2.87 & 41.66 & 48.55 & 56.09 & 45.94 & 50.19 $\pm$ 4.30 \\
UNEG~~\cite{gonzalez2020wrong}         & $\dag$~$*$ & \cmark & \xmark & 72.45 & 59.36 & 56.31 & 50.57 & 55.41 $\pm$ 3.64 & 56.34 & 61.57 & 60.47 & 47.59 & 56.54 $\pm$ 4.13 \\
Lifelong-nnUNet~~\cite{gonzalez2023lifelong} & $\dag$~$*$ & \cmark & \xmark & 77.41 & 63.96 & 50.25 & \textbf{65.92} & 60.04 $\pm$ 4.15 & 59.63 & 63.27 & 52.53 & 57.74 & 57.84 $\pm$ 5.68 \\
CFIL~~\cite{chen2024classifier}        & $\dag$     & \cmark & \cmark & 69.43 & 57.68 & 50.44 & 60.15 & 56.09 $\pm$ 4.12 & 50.39 & 58.02 & 51.27 & 53.97 & 54.42 $\pm$ 2.77 \\
VIL~~\cite{park2024versatile}          & $\dag$~$*$ & \xmark & \xmark & 56.32 & 41.63 & 36.87 & 40.59 & 39.70 $\pm$ 1.42 & 21.59 & 43.77 & 38.32 & 41.09 & 41.06 $\pm$ 1.62 \\\hline
FGCARL + nnU-Net                        & $\dag$~$*$ & \cmark & \xmark & \underline{78.41} & 64.17 & 51.43 & \underline{65.66} & 60.42 $\pm$ 6.56 & \underline{67.19} & 64.79 & 59.94 & \underline{58.14} & \underline{60.96} $\pm$ 6.31 \\
FGCARL + Swin UNETR           & $\dag$~$*$ & \cmark & \cmark & \textbf{80.13} & \textbf{73.66} & \textbf{62.85} & 63.02 & \textbf{66.51} $\pm$ 5.06 & \textbf{70.63} & \textbf{73.79} & \textbf{70.37} & \textbf{60.67} & \textbf{68.28} $\pm$ 5.56 \\
\bottomrule
\end{tabular}} 
\label{tab1}  
\end{table*}

\begin{figure*}
    \centering
    \includegraphics[width=\linewidth]{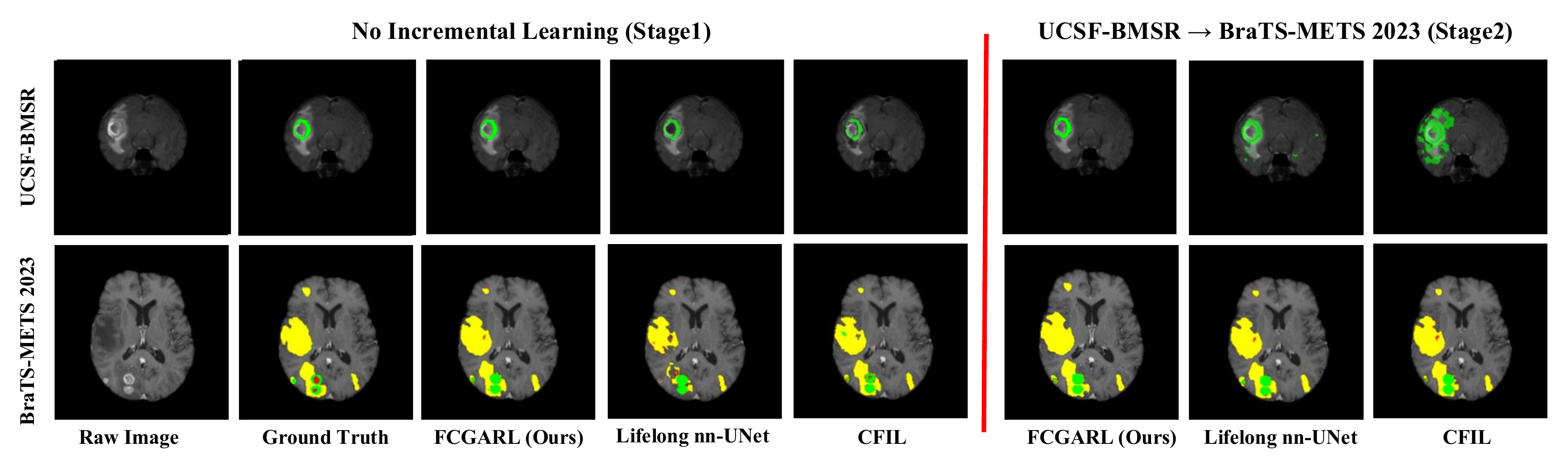} 
    \caption{Qualitative comparison of FGCARL with representative state-of-the-art incremental learning methods on the original and incrementally adapted domains. Examples are shown for the UCSF-BMSR $\rightarrow$ BraTS-METS 2023 adaptation setting. For the previously learned UCSF-BMSR domain, the predictions illustrate the ability of each method to retain previously acquired segmentation knowledge after incremental adaptation. For the BraTS-METS 2023 domain, the predictions demonstrate adaptation to the new domain and expanded label space, including ET, TC, and SNFH.}
    \label{fig4}
\end{figure*}

\textbf{Overall segmentation performance.} FGCARL consistently achieved superior performance under the unified domain- and class-incremental setting (Table~\ref{tab1}). With Swin UNETR, FGCARL achieved a mean DSC of $68.28\%$ on the target BraTS-METS 2023 domain, substantially outperforming the strongest competing method, CLAMTS, which achieved $60.94\%$. This corresponds to an improvement of $7.34\%$ points. FGCARL also achieved the best performance for individual target structures, obtaining DSC scores of $73.79\%$, $70.37\%$, and $60.67\%$ for ET, TC, and SNFH, respectively. These results indicated that FGCARL effectively adapted to both newly introduced anatomical/pathological classes and substantial changes in the imaging distribution. The advantage of FGCARL was also evident when using a CNN-based segmentation architecture, nnU-Net. The employment of FGCARL to nnU-Net achieved a mean DSC of $60.96\%$ on BraTS-METS 2023, outperforming the strongest competing methods under the same incremental setting. Notably, FGCARL + nnU-Net achieved a $67.19\%$ DSC for the previously learned UCSF-BMSR domain, demonstrating that the proposed framework is not restricted to transformer-based architectures. The consistent improvements obtained with both nnU-Net and Swin UNETR support the architecture-agnostic design of FGCARL.

\textbf{Mitigation of catastrophic forgetting.} In addition to improving target-domain adaptation, FGCARL substantially preserved knowledge acquired from the source domain. Before incremental adaptation, FGCARL + Swin UNETR achieved $80.13\%$ mean DSC on UCSF-BMSR, whereas after adaptation to BraTS-METS 2023, it retained a DSC of $70.63\%$. Similarly, FGCARL + nnU-Net retained $67.19\%$ DSC after adaptation compared with $78.41\%$ before incremental training. Although some performance degradation is expected when the model is optimized for a substantially different domain and expanded label space, FGCARL maintained considerably stronger source-domain performance than competing methods. For example, on UCSF-BMSR after adaptation, CLAMTS and Lifelong-nnUNet achieved substantially lower performance than FGCARL. These results demonstrate the effectiveness of the FGKR strategy in protecting parameters associated with previously acquired knowledge and mitigating catastrophic forgetting.

\textbf{Robustness to simultaneous domain and class shifts.} Conventional CIL methods generally assume that the imaging distribution remains relatively stable and focus primarily on learning newly introduced classes, whereas DIL methods primarily address changes in data distributions. In contrast, the Stage 2 setting requires the model to simultaneously adapt to a new institution/dataset, different modality configurations, and an expanded label space. Most existing methods exhibited a substantial performance degradation under this combined shift. FGCARL consistently maintained high segmentation performance across both previously learned and newly introduced targets, demonstrating that the proposed combination of MAIR and FGKR provides a unified mechanism for feature adaptation and knowledge retention.

\textbf{Qualitative comparison.} Figure~\ref{fig4} presents qualitative comparisons between FGCARL and representative incremental learning methods, including Lifelong-nnUNet and CFIL, on both the original and incrementally adapted domains. On the UCSF-BMSR samples, which represent the previously learned domain, FGCARL preserves the overall morphology and spatial localization of the metastatic lesions after adaptation to BraTS-METS 2023. In contrast, Lifelong-nnUNet and CFIL exhibit more pronounced prediction errors, including incomplete lesion delineation, spurious predictions, and deviations from the reference boundaries. These observations are consistent with the quantitative results and indicate that FGCARL better preserves previously acquired segmentation knowledge during incremental adaptation. On BraTS-METS 2023, FGCARL produces segmentation masks that more closely follow the reference annotations, particularly in delineating the spatial extent and boundaries of different tumor subregions. The competing methods exhibit greater segmentation discrepancies, including missing regions and inaccurate predictions of individual tumor classes. Notably, FGCARL maintains relatively coherent predictions for ET, TC, and SNFH despite the simultaneous changes in imaging distribution and label space. Overall, the qualitative results demonstrate that FGCARL achieves a favorable balance between plasticity, i.e., learning new domain and class knowledge, and stability, i.e., retaining previously acquired knowledge.

Overall, the results demonstrate that FGCARL provides a more effective solution for unified domain- and class-incremental medical image segmentation than existing SOTA methods. Its consistent performance across both Swin UNETR and nnU-Net, together with its superior target-domain adaptation and source-domain knowledge retention, validates the effectiveness, generalizability, and architecture-agnostic nature of the proposed framework.

\subsection{Ablation Study}
\subsubsection{The effectiveness of MAIR}
We first evaluated the effectiveness of the proposed MAIR for domain-incremental learning under unaligned modality settings. We compared MAIR with representative SOTA DIL methods and several normalization-based approaches, including Batch Renormalization and Continuous Batch Renormalization. We also compared the incorporation of MAIR with the only forward MAIR. Experiments were conducted using both nnU-Net and Swin UNETR, with domain adaptation evaluated in both directions: UCSF-BMSR $\rightarrow$ BrainMetShare and BrainMetShare $\rightarrow$ UCSF-BMSR. This bidirectional evaluation allows us to assess whether MAIR provides consistent adaptation benefits across different source-target domain configurations.

\begin{table*}[!t]
\centering
\caption{Performance comparison of MAIR with SOTA domain-incremental learning methods and normalization-based approaches under unaligned modality settings. Models are evaluated on UCSF-BMSR and BrainMetShare in both adaptation directions. Stage 1 reports performance without incremental adaptation, while Stage 2 reports performance after adaptation between the two domains. \textbf{Bold} and \underline{underline} indicate the best and second-best results, respectively.}
\renewcommand{\arraystretch}{1.2}
\scriptsize
\resizebox{\textwidth}{!}{%
\begin{tabular}{@{}l|cc|cc|cc@{}}
\toprule
\multirow{2}{*}{Methods} & \multicolumn{2}{c|}{No Incremental Learning (Stage 1)} & \multicolumn{2}{c|}{UCSF-BMSR \textrightarrow BrainMetShare (Stage 2)} & \multicolumn{2}{c}{BrainMetShare \textrightarrow UCSF-BMSR (Stage 2)} \\
\cmidrule{2-7}
& UCSF-BMSR & BrainMetShare & UCSF-BMSR & BrainMetShare & UCSF-BMSR & BrainMetShare \\
\midrule
ProCA~\cite{jiang2022prototypical}          & 56.25 & 41.38 & 40.41 & 42.69 & 57.61 & 24.85 \\
EWC~\cite{kirkpatrick2017overcoming}        & 55.51 & 42.99 & 38.42 & 44.67 & 56.88 & 25.41 \\
MAS-LR~\cite{ozgun2020importance}           & 55.38 & 43.15 & 41.08 & 47.92 & 57.21 & 28.89 \\
Msmt-medseg~\cite{tian2022multi}            & 71.61 & 58.43 & 53.47 & 60.01 & 72.13 & 40.64 \\
MDIL-SS~\cite{garg2022multi}                & 53.38 & 39.02 & 36.93 & 40.19 & 54.19 & 24.57 \\
HSI~\cite{liu2023incremental}               & 65.46 & 52.73 & 49.73 & 53.58 & 66.27 & 32.54 \\
CGR~\cite{li2024comprehensive}              & 54.63 & 43.31 & 41.44 & 45.17 & 55.04 & 27.82 \\
Style-Oriented~\cite{li2022domain}          & 57.29 & 45.59 & 43.79 & 46.57 & 58.94 & 30.22 \\
Memory Replay~\cite{bera2023memory}         & 60.45 & 49.88 & 56.75 & 46.91 & 61.73 & 41.60 \\
UNEG~\cite{gonzalez2020wrong}               & 72.45 & 59.47 & 66.75 & 62.38 & 73.07 & 49.39 \\
Lifelong nnUNet~\cite{gonzalez2023lifelong} & 77.41 & 63.96 & 69.89 & 65.11 & 78.68 & 54.44 \\
CFIL~\cite{chen2024classifier}              & 69.43 & 57.91 & 65.47 & 59.01 & 71.72 & 50.47 \\
Mdvit~\cite{du2023mdvit}                    & 73.59 & 56.67 & 60.57 & 61.94 & 77.19 & 52.57 \\
\midrule
nnU-Net                                     & 77.41 & 63.96 & 45.42 & 65.74 & 77.63 & 44.11 \\
nnU-Net + Batch Renormalization             & 77.41 & 63.96 & 65.71 & 64.48 & 77.69 & 44.40 \\
nnU-Net + Continuous Batch Renormalization  & 77.41 & 63.96 & 68.44 & 65.97 & 77.91 & 52.88 \\ 
nnU-Net + Forward MAIR                       & 77.49 & 62.14 & 50.13 & 65.33 & 77.98 & 47.99 \\
nnU-Net + Forward $\&$ Backward MAIR         & \underline{77.93} & \underline{64.11} & \underline{72.47} & \underline{66.15} & \underline{78.96} & \underline{56.27} \\
\midrule
Swin UNETR                                  & 74.68 & 63.31 & 40.31 & 66.29 & 75.93 & 41.08 \\
Swin UNETR + Batch Renormalization          & 63.51 & 45.19 & 56.98 & 47.77 & 64.09 & 39.51 \\
Swin UNETR + Continuous Batch Renormalization & 63.51 & 45.19 & 60.85 & 51.55 & 66.29 & 42.88 \\ 
Swin UNETR + Forward MAIR                    & 72.14 & 60.27 & 39.55 & 67.60 & 73.59 & 38.57 \\
Swin UNETR + Forward $\&$ Backward MAIR      & \textbf{80.13} & \textbf{65.63} & \textbf{77.32} & \textbf{68.27} & \textbf{81.62} & \textbf{62.15} \\
\bottomrule
\end{tabular}%
} 
\label{tab2}
\end{table*}

\begin{figure*}[!ht]
    \centering
    \includegraphics[width=\linewidth]{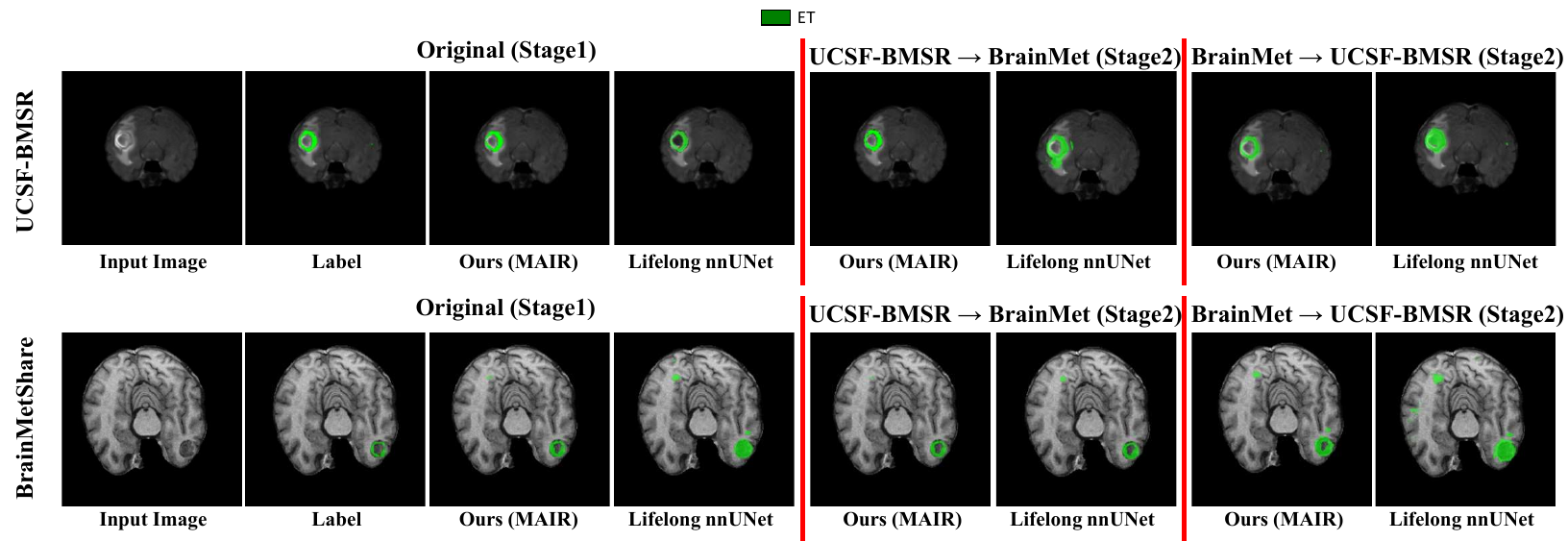}  
    \caption{Qualitative comparison of MAIR with Lifelong nnU-Net on the original (Stage 1) and incrementally adapted domains (UCSF-BMSR \textrightarrow BrainMetShare, BrainMetShare \textrightarrow UCSF-BMSR). The highlighted regions (green, ET) represent effective tumor boundary segmentation, demonstrating MAIR's robustness in handling domain shifts and incremental tasks.}     
    \label{fig5}
\end{figure*}

\textbf{Comparison with existing DIL methods.} MAIR consistently achieved the strongest performance among the evaluated methods across both segmentation architectures and adaptation directions (Table~\ref{tab2}). With Swin UNETR, the proposed Forward $\&$ Backward MAIR achieved DSC scores of $77.32\%$ and $68.27\%$ on UCSF-BMSR and BrainMetShare, respectively, when adapting from UCSF-BMSR to BrainMetShare. In the reverse adaptation setting, it achieved $81.62\%$ and $62.15\%$ DSC on UCSF-BMSR and BrainMetShare, respectively. These results substantially outperformed existing DIL methods, including Lifelong nnUNet, UNEG, Mdvit, and CFIL. In particular, for UCSF-BMSR $\rightarrow$ BrainMetShare, Forward $\&$ Backward MAIR improved the BrainMetShare DSC from $65.11\%$ achieved by Lifelong nnUNet to $68.27\%$, while simultaneously retaining substantially higher performance on the previously learned UCSF-BMSR domain ($77.32\%$).

\textbf{Effect of bidirectional MAIR.} The ablation results demonstrate that the the combination of Forward and Backward components is critical to the effectiveness of MAIR. For nnU-Net, Forward MAIR alone achieved only $50.13\%$ DSC on UCSF-BMSR after adaptation from UCSF-BMSR to BrainMetShare, whereas incorporating both Forward and Backward MAIR increased the DSC to $72.47\%$. A similar improvement was observed for the reverse adaptation direction, where the UCSF-BMSR DSC increased from $77.98\%$ with Forward MAIR alone to $78.96\%$ with Forward $\&$ Backward MAIR, while the BrainMetShare DSC increased from $47.99\% $to $56.27\%$. These results indicate that forward normalization alone can facilitate feature adaptation to the incoming domain but is insufficient to preserve the feature statistics associated with previously learned domains. The backward alignment mechanism provides the complementary capability of restoring domain-specific feature characteristics, thereby improving the stability of previously acquired representations during incremental adaptation.

\textbf{Comparison with conventional normalization strategies.} Forward $\&$ Backward MAIR also consistently outperformed Batch Renormalization and Continuous Batch Renormalization. For nnU-Net under UCSF-BMSR $\rightarrow$ BrainMetShare adaptation, Continuous Batch Renormalization achieved $68.44\%$ and $65.97\%$ DSC on UCSF-BMSR and BrainMetShare, respectively, whereas Forward $\&$ Backward MAIR achieved $72.47\%$ and $66.15\%$. The advantage was more pronounced for Swin UNETR, where Forward $\&$ Backward MAIR achieved $77.32\%$ and $68.27\%$, compared with $60.85\%$ and $51.55\%$ obtained using Continuous Batch Renormalization. These results suggest that simply updating normalization statistics is insufficient for handling substantial cross-domain distribution shifts, particularly for transformer-based architectures. In contrast, the bidirectional normalization and denormalization mechanism of MAIR provides more effective feature adaptation while maintaining previously learned representations.

\textbf{Generalization across segmentation architectures.} The consistent improvements obtained with both nnU-Net and Swin UNETR further demonstrate that MAIR is not tied to a particular segmentation architecture. Notably, the improvement was particularly substantial for Swin UNETR, where conventional Batch Renormalization and Continuous Batch Renormalization substantially degraded performance relative to the original model, whereas Forward $\&$ Backward MAIR improved both source- and target-domain performance. Without incremental adaptation, Swin UNETR achieves $74.68\%$ and $63.31\%$ DSC on UCSF-BMSR and BrainMetShare, respectively; after incorporating Forward $\&$ Backward MAIR, these values increased to $80.13\%$ and $65.63\%$, respectively. This observation supports the ability of MAIR to provide stable feature representations under domain shifts while remaining compatible with both CNN- and ViT-based segmentation architectures.

\textbf{Qualitative comparison.} Figure~\ref{fig5} presents qualitative comparisons between FGCARL and Lifelong-nnUNet on both the original and incrementally adapted domains. In the previously learned domain, FGCARL preserved the overall morphology and spatial localization of the metastatic lesions after adaptation to the new domain. In contrast, Lifelong-nnUNet exhibit more pronounced prediction errors, including incomplete lesion delineation, spurious predictions, and deviations from the reference boundaries.

Overall, these experiments demonstrate that the effectiveness of MAIR arises from the combination of its forward and backward operations rather than from normalization alone. Forward MAIR facilitates the stabilization of incoming domain representations, while Backward MAIR restores domain-specific feature characteristics and supports retention of previously learned representations. Their combination enables robust adaptation in both directions and across distinct segmentation architectures, providing an effective foundation for domain-incremental medical image segmentation.

\subsubsection{The effectiveness of FGKR}
To further evaluate the effectiveness of FGKR under class-incremental learning, we conduct experiments on BTCV and LiTS using a stage-wise incremental learning setting, where new classes are introduced sequentially across domains. All methods achieved identical performance at Stage 1, providing a consistent starting point for evaluating their ability to adapt to the subsequent incremental stage (Table~\ref{tab3}).

\begin{table*}[t]
\centering
\caption{Ablation study of FGKR under class-incremental learning. Results are averaged across all classes, with DSC ($\%$) further averaged across the corresponding sub-regions. Stage 1 represents the initial non-incremental learning setting, while Stage 2 denotes incremental adaptation from LiTS to BTCV. \textbf{Bold} and \underline{underline} indicate the best and second-best results, respectively.}
\resizebox{0.8\textwidth}{!}{
\begin{tabular}{l|c|c|c|c}
\toprule
\multirow{2}{*}{Methods} & \multicolumn{2}{c|}{No Incremental Learning (Stage 1)} & \multicolumn{2}{c}{BTCV \textrightarrow LiTs (Stage 2)}  \\
\cline{2-5}
& BTCV & LiTs & BTCV & LiTs \\
\midrule
LwF                   & 88.90$\pm$1.33 & 43.17$\pm$5.38 & 77.02$\pm$2.48 & 45.65$\pm$4.43  \\
EWC                   & 88.90$\pm$1.33 & 43.17$\pm$5.38 & 79.04$\pm$2.28 & 45.94$\pm$4.18  \\
Synaptic Intelligence & 88.90$\pm$1.33 & 43.17$\pm$5.38 & 79.61$\pm$3.39 & 45.32$\pm$6.01  \\
CWR                   & 88.90$\pm$1.33 & 43.17$\pm$5.38  & \underline{80.24}$\pm$1.65 & 46.30$\pm$3.97  \\
\hline
FGKR + nnUNet         & 88.90$\pm$1.33 & 43.17$\pm$5.38 & 79.67$\pm$2.71 & \underline{46.48}$\pm$5.08 \\
FGKR + Swin UNETR     & 88.90$\pm$1.33 & 43.17$\pm$5.38 & \textbf{83.08}$\pm$1.58 & \textbf{47.59}$\pm$4.12   \\
\bottomrule
\end{tabular}}
\label{tab3}
\end{table*}

After introducing the new classes in Stage 2, FGKR consistently achieved competitive or superior performance compared with existing class-incremental learning methods. In particular, FGKR combined with Swin UNETR achieved the best performance on both BTCV and LiTS, reaching $83.08\% $and $47.59\%$ DSC, respectively. On BTCV, this corresponded to a substantial improvement over the strongest competing baseline, demonstrating the effectiveness of FGKR in preserving and adapting semantic knowledge when new classes are introduced. FGKR also achieved the best performance on LiTS when paired with either nnUNet or Swin UNETR, with the latter providing the strongest overall results.

The consistent improvements across the two datasets and segmentation architectures demonstrate that the effectiveness of FGKR is not restricted to a particular network or dataset. Notably, the stronger performance obtained with Swin UNETR suggests that FGKR can effectively complement different segmentation architectures and facilitate knowledge retention and adaptation during class-incremental learning. Overall, these results provide further evidence that FGKR is robust to sequential class expansion and domain variation in realistic incremental segmentation scenarios.

\subsection{The analysis on MAIR configurations} 
We further investigate the effectiveness of MAIR by incorporating it into two representative segmentation architectures, nnU-Net and Swin UNETR, and evaluating their performance on UCSF-BMSR, BrainMetShare, and BraTS-METS 2023. MAIR consistently improves segmentation performance across both architectures, with more pronounced gains observed for Swin UNETR (Table~\ref{tab4}).

For nnU-Net, incorporating MAIR improves performance across all three datasets, and applying MAIR to all layers consistently outperforms the single-layer configuration. On BraTS-METS 2023, for example, the mean DSC across TC, SNFH, and ET increases from $61.34\%$ for the original nnU-Net to $64.14\%$ with all-layer MAIR. These results demonstrate that MAIR can effectively enhance conventional convolutional segmentation networks and that its benefits are further strengthened when integrated throughout the network.

The improvement is more pronounced for Swin UNETR. The baseline model achieves a mean DSC of $63.15\%$ across the three BraTS-METS 2023 tumor regions, which increases to $67.57\%$ with all-layer MAIR. Moreover, the performance consistently improves as MAIR is incorporated into more layers, from single-layer to two-layer and finally all-layer integration. The all-layer configuration achieves the best performance across the three tumor regions and also obtains the highest DSC on both UCSF-BMSR and BrainMetShare. These results suggest that distributing MAIR throughout the network enables Swin UNETR to better accommodate variations in data distributions across different datasets.

We further compare MAIR with alternative normalization strategies. Replacing the original normalization with batch normalization (BN) or instance normalization (IN) substantially degrades performance, whereas MAIR consistently improves upon the original Swin UNETR. In particular, BN and IN achieve mean DSCs of only $56.02\%$ and $59.31\%$, respectively, compared with $67.57\%$ for all-layer MAIR. This comparison indicates that the performance gains are not simply attributable to changing the normalization strategy. Instead, the progressive improvement from single-layer to all-layer MAIR suggests that integrating MAIR throughout the network is important for effectively mitigating distribution shifts and improving segmentation robustness.

Overall, these results demonstrate that MAIR is compatible with both convolutional and transformer-based segmentation architectures and provides consistent performance improvements under heterogeneous data distributions. The superior performance of the all-layer configuration further supports the effectiveness of applying MAIR across multiple normalization layers rather than restricting it to a single layer.

\begin{table*}[!t]
\centering
\caption{Performance comparison of different MAIR configurations incorporated into nnU-Net and Swin UNETR on the UCSF-BMSR, BrainMetShare, and BraTS-METS 2023 datasets. DSC ($\%$) is reported for the indicated tumor regions. For BraTS-METS 2023, TC, SNFH, and ET denote the corresponding tumor subregions, and the mean $\pm$ SD is computed across the three regions. \textbf{Bold} and \underline{underline} indicate the best and second-best results, respectively.}
\resizebox{1.0\textwidth}{!}{
\begin{tabular}{l|c|c|c|c|c|c}
\toprule
\multirow{2}{*}{Methods} & UCSF-BMSR & BrainMetShare & \multicolumn{4}{c}{BraTS-METS 2023} \\
\cline{2-7}
& ET & ET & TC & SNFH & ET & Mean $\pm$ SD \\
\midrule
3D U-Net      \cite{cciccek20163d}       & 72.45 & 59.47 & 49.36 & 56.31 & 50.57 & 52.08 $\pm$ 3.03 \\
3D TransUNet \cite{chen2024transunet}    & 79.35 & 64.44 & 52.31 & 67.11 & 71.43 & 63.62 $\pm$ 8.19 \\
V-Net        \cite{milletari2016v}       & 73.56 & 62.99 & 56.12 & 59.72 & 60.14 & 58.66 $\pm$ 1.80 \\
UNETR        \cite{hatamizadeh2022unetr} & 72.72 & 60.38 & 62.35 & 57.51 & 56.93 & 58.93 $\pm$ 2.43 \\
nnFormer     \cite{zhou2023nnformer}     & 73.98 & 61.73 & 48.34 & 64.17 & 67.73 & 60.08 $\pm$ 8.43 \\
\midrule
nnU-Net                                  & 77.41 & 63.96 & 50.25 & 65.92 & 67.86 & 61.34 $\pm$ 7.88 \\
nnU-Net + MAIR (Single Layer)            & 78.15 & 64.41 & 54.63 & \underline{65.71} & \underline{68.59} & 62.98 $\pm$ 6.02 \\
nnU-Net + MAIR (All Layers)              & 79.07 & 64.87 & 57.28 & \textbf{66.21} & \textbf{68.92} & 64.14 $\pm$ 4.97 \\
\midrule
Swin UNETR                               & 74.68 & 63.31 & 69.35 & 60.62 & 59.48 & 63.15 $\pm$ 4.41 \\
Swin UNETR + BN                          & 63.51 & 52.10 & 59.55 & 51.26 & 57.25 & 56.02 $\pm$ 3.49 \\
Swin UNETR + IN                          & 65.41 & 56.34 & 62.17 & 56.52 & 59.25 & 59.31 $\pm$ 2.31 \\
Swin UNETR + MAIR (Single Layer)         & 80.13 & 65.63 & 73.66 & 62.85 & 63.02 & 66.51 $\pm$ 5.06 \\
Swin UNETR + MAIR (2 Layers)             & \underline{80.65} & \underline{65.91} & \underline{73.79} & 62.88 & 64.11 & \underline{66.93} $\pm$ 4.88 \\
Swin UNETR + MAIR (All Layers)           & \textbf{81.34} & \textbf{66.42} & \textbf{74.48} & 63.18 & 65.05 & \textbf{67.57} $\pm$ 4.95 \\
\bottomrule
\end{tabular}
}
\label{tab4}
\end{table*}

\subsection{Generalization of MAIR on 2D medical images}
To further evaluate the generalizability of MAIR beyond the primary experimental setting, we conducted experiments on the M$\&$Ms dataset, which comprised cardiac MRI data acquired from multiple vendors. MAIR consistently improved segmentation performance across all four vendors, demonstrating its robustness to vendor-induced domain shifts (Table~\ref{tab5}). When combined with nnU-Net, MAIR outperformed all competing continual domain adaptation methods across vendors, with particularly notable gains over the strongest baseline on Vendor C.

We further combined MAIR with Swin UNETR to examine its compatibility with different segmentation backbones. MAIR + Swin UNETR achieved the best performance on three of the four vendors and remained highly competitive on the remaining vendor. The consistent performance of MAIR with both nnU-Net and Swin UNETR demonstrate that its effectiveness is not dependent on a specific segmentation architecture. Overall, these results highlight the strong generalization capability of MAIR across heterogeneous vendors and different backbone architectures for 2D medical image segmentation.

\begin{table}[t]
\centering
\caption{Evaluation of the generalization of MAIR across vendors and segmentation backbones on the M$\&$Ms dataset. DSC ($\%$) is reported for each vendor. \textbf{Bold} and \underline{underline} indicate the best and second-best results, respectively.} 
\resizebox{\columnwidth}{!}{
\begin{tabular}{l|cccc}
\toprule
Methods & Vendor A & Vendor B & Vendor C & Vendor D \\
\midrule
MAS-LR~~\cite{ozgun2020importance}     & 84.89 & 88.36 & 86.57 & 84.62 \\
LwF~~\cite{li2017learning}             & 85.42 & 88.78 & 86.22 & 85.14 \\
EWC~~\cite{kirkpatrick2017overcoming}  & 84.74 & 88.86 & 86.79 & 85.33 \\
GR~~\cite{shin2017continual}           & 85.11 & 89.04 & 86.44 & 85.24 \\
Style-Oriented~~\cite{li2022domain}    & 85.91 & 89.12 & 86.85 & 85.44 \\
\hline
MAIR + nnU-Net                         & \underline{87.57} & \underline{89.41} & \textbf{87.69} & \underline{86.04} \\
MAIR + Swin UNETR                      & \textbf{87.95} & \textbf{89.73} & \underline{87.51} & \textbf{86.17} \\
\bottomrule
\end{tabular}}
\label{tab5}
\end{table}

\subsection{Generalization across network architecture}
To evaluate the architecture generalization of FGCARL, we incorporated FGCARL into two representative 3D segmentation architectures, nnU-Net and Swin UNETR, and evaluated their performance under the UCSF-BMSR $\rightarrow$ BraTS-METS 2023 incremental learning setting (Table~\ref{tab6}). FGCARL can be directly integrated into different network architectures as an adaptive normalization mechanism, without requiring architecture-specific modifications.

FGCARL consistently improved segmentation performance for both architectures. In Stage 1, where the networks were trained and evaluated within the same domain, incorporating FGCARL improved the baseline performance of both nnU-Net and Swin UNETR on UCSF-BMSR and BraTS-METS 2023. Notably, the improvement is particularly pronounced for Swin UNETR, indicating that the adaptive normalization provided by FGCARL can also enhance the representation learned by transformer-based segmentation networks even in the absence of domain shifts.

More importantly, FGCARL maintained its effectiveness during continual adaptation from UCSF-BMSR to BraTS-METS 2023. In Stage 2, FGCARL achieved the highest performance among the evaluated incremental learning methods on both the source and target domains for both network architectures. For nnU-Net, FGCARL substantially improved retention of the previously learned UCSF-BMSR knowledge while simultaneously achieving the strongest performance on BraTS-METS 2023. A similar trend was observed with Swin UNETR, where FGCARL considerably outperformed the competing incremental learning methods on both domains. These results demonstrate that the benefits of FGCARL are not tied to a particular segmentation architecture and that its adaptive normalization and knowledge retention mechanisms can be effectively combined with both convolutional and transformer-based networks.

\begin{table*}[!t]
\caption{Performance comparison of FGCARL and other incremental learning methods using nnU-Net and Swin UNETR for 3D multi-modal medical image segmentation. Stage 1 evaluates in-domain performance without incremental learning, whereas Stage 2 evaluates continual adaptation from UCSF-BMSR to BraTS-METS 2023. DSC ($\%$) is reported for single-class segmentation and as mean $\pm$ standard deviation for multi-class segmentation. \textbf{Bold} and \underline{underline} indicate the best and second-best results, respectively.}
\centering
\resizebox{\textwidth}{!}{
\begin{tabular}{l|l|c|c|c|c}
\toprule
\multirow{2}{*}{Networks} & \multirow{2}{*}{Incremental Methods} & \multicolumn{2}{c|}{No Incremental Learning (Stage 1)} & \multicolumn{2}{c}{UCSF-BMSR $\rightarrow$ BraTS-METS 2023 (Stage 2)} \\
\cline{3-6}
& & UCSF-BMSR & BraTS-METS 2023 & UCSF-BMSR & BraTS-METS 2023  \\
\hline
\multirow{5}{*}{nnU-Net} 
& CWR + Batch Renormalization & 77.41 & 60.04$\pm$4.15 & 58.18 & 60.21$\pm$4.59 \\
& CoFit                       & 77.41 & 60.04$\pm$4.15 & 56.94 & 59.71$\pm$4.01 \\
& Lifelong                    & 77.41 & 60.04$\pm$4.15 & 59.63 & 57.84$\pm$2.77 \\
& FGCARL                      & \underline{78.41} & \underline{60.42}$\pm$6.56 & \underline{67.19} & \underline{60.96}$\pm$6.31 \\
\hline
\multirow{4}{*}{Swin UNETR} 
& CWR + Batch Renormalization & 63.51 & 49.36$\pm$6.57 & 60.46 & 51.48$\pm$7.16  \\
& CoFit                       & 63.51 & 49.36$\pm$6.57 & 62.61 & 51.39$\pm$6.97  \\
& FGCARL                      & \textbf{80.13} & \textbf{66.51}$\pm$5.06 & \textbf{70.63} & \textbf{68.28}$\pm$5.56   \\
\bottomrule
\end{tabular}
}
\label{tab6}
\end{table*}

\begin{figure*}
\centering
\includegraphics[width=0.8\textwidth]{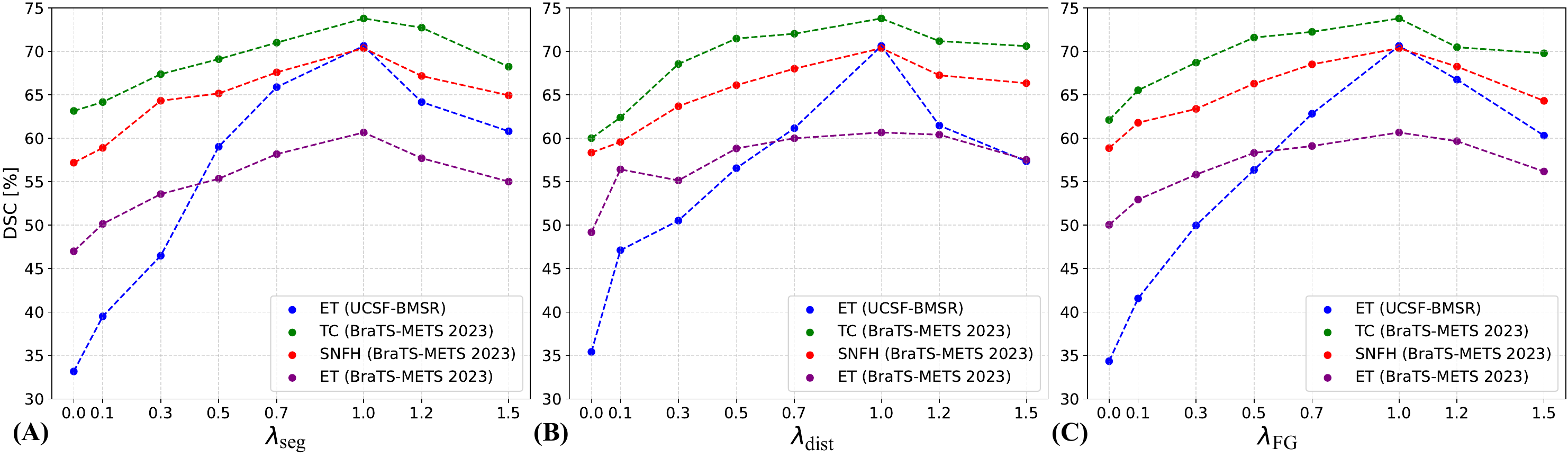}
\caption{Parameter sensitivity analysis of loss regularization weights by evaluating DSC scores in the UCSF-BMSR $\rightarrow$ BraTS-METS 2023 setting. Impact of varying (A) $\lambda_{seg}$, (B) $\lambda_{dist}$, (C) $\lambda_{FG}$ across evaluation subregions in the BraTS-METS 2023.}
\label{fig6}
\end{figure*}

\subsection{Parameter sensitivity analysis}
We evaluated the sensitivity of segmentation performance, measured by DSC, to the regularization weighting hyperparameters in the UCSF-BMSR $\rightarrow$ BraTS-METS 2023 setting. Specifically, we independently varied the weights of the three regularization components, $\lambda_{\mathrm{seg}}$, $\lambda_{\mathrm{dist}}$, and $\lambda_{\mathrm{FG}}$ (Figure~\ref{fig6} (A)–(C)), respectively.

Overall, the three regularization terms exhibit similar trends across the evaluated tumor subregions. Increasing the regularization weight from $\lambda=0$ generally improves segmentation performance, indicating that each component contributes to effective continual adaptation and knowledge retention. Across the evaluated components and tumor subregions, $\lambda=1.0$ provides a favorable balance between retaining previously learned knowledge and adapting to the target domain, yielding consistently strong performance on both the source and target datasets. We therefore set the regularization weight to $\lambda=1.0$ in the remaining experiments.

\section{Conclusion}
In this work, we presented Fisher-Guided Continuously Adaptive Representation Learning (FGCARL), a unified framework for continual 3D medical image segmentation under evolving domain and class shifts. By integrating Mutual Adaptive Instance Representation (MAIR) for adaptive domain alignment with Fisher-Guided Knowledge Retention (FGKR) for selective preservation of previously acquired knowledge, FGCARL enables continuous adaptation while mitigating catastrophic forgetting. Extensive experiments across multiple cross-institutional segmentation tasks and diverse segmentation architectures demonstrated that FGCARL consistently outperforms existing continual learning approaches, achieving effective adaptation to newly encountered domains and classes while maintaining knowledge of previously learned ones. These results highlight the potential of FGCARL as a unified and generalizable solution for continual medical image segmentation in dynamic clinical environments.

\section*{Declaration of competing interests}
The authors declare that they have no known competing financial interests or personal relationships that could have appeared to influence the work reported in this paper.

\section*{CRediT authorship contribution statement}
XY: conceptualization, methodology, writing the original draft, reviewing, and editing; JY: conceptualization, methodology, writing the original draft, reviewing, and editing, visualization; PQ: conceptualization, writing the original draft, reviewing, and editing; WM: reviewing, and editing; XL: reviewing, and editing.








\bibliographystyle{cas-model2-names}

\bibliography{reference}



\end{document}